%% file: dmpi_0926_rr.tex
\documentclass[12pt,authoryear]{elsarticle}

\usepackage{lineno,hyperref}

\modulolinenumbers[5]

\usepackage{graphicx}

\usepackage{setspace}
\usepackage{adjustbox}
\usepackage{array}
\usepackage[font=small]{caption}

\usepackage{tgtermes}
\usepackage[T1]{fontenc}

\usepackage{amssymb}
\usepackage{amsmath}
\usepackage{amsthm}
\usepackage{enumerate}
\usepackage{lscape}
\usepackage{ascmac}
\usepackage[font=small]{caption}

\usepackage{float}
\usepackage[font=large, labelfont=bf]{caption}

\usepackage[top=1in, bottom=1in, left=1in, right=1in]{geometry}

\usepackage[most]{tcolorbox}
\tcbset{
  colback=white,       % 背景色
  colframe=black,      % 枠線
  arc=0mm,             % 角の丸み
  left=2mm, right=2mm, boxsep=2pt
}

\usepackage{tikz}
\usetikzlibrary{positioning, calc, shapes, arrows.meta}

\biboptions{authoryear}

\begin{document}

\setcounter{page}{0}

\pagestyle{empty}

\vspace*{-0.5in}

\begin{center}
\textsc{{\Large \textbf{Distribution-Matching Posterior Inference for}}\\
\vspace{1.0mm} {\Large \textbf{Incomplete Structural Models}}}

\vspace{0.25in}

\normalsize

\begin{tabular}{c}
{\large Takashi Kano}   \\
Graduate School of Economics  \\
Hitotsubashi University         \\
Naka 2-1, Kunitachi, Tokyo  \\
186-8601, JAPAN     \\
Email: tkano@econ.hit-u.ac.jp  \\
\end{tabular}
\end{center}

\vspace{0.05in}

\begin{center}

\begin{center}
Revised: October 23, 2026 \\
%\textit{Incomplete and Preliminary}
\end{center}

\end{center}

\noindent
\textit{Abstract}\hspace{2mm}$\overline{\hspace{6.32in}}$\newline
Distribution-matching posterior inference (DMPI) extends the minimal econometric interpretation to generalized posterior inference and model comparison for incomplete structural models. It combines a structural-parameter prior with exact Dirichlet--multinomial scores comparing empirical and model-implied distributions of selected population moments. Ensemble size controls structural discipline, and annealed sequential Monte Carlo computes posterior and generalized evidence. In New Keynesian Phillips Curve simulations, a stochastically singular model remains useful under moderate discipline despite failing as an exact representation. In a U.S. three-equation New Keynesian DSGE application, restoring full rank raises or lowers generalized evidence depending on the structural innovation added.
\hspace{0.5in}
\newline $\overline{\hspace{7in}}$

\vspace{3mm}

\vspace{1mm}


\vspace{1mm}


\vspace{1.5in}

\footnotesize
\noindent
$^{\dag}$ I thank Editor Sylvia Fr\"uhwirth-Schnatter for her constructive comments and guidance during the revision process. I also thank Gianni Amisano, Toru Kitagawa, Jim Nason, Aureo de Paula, Francisco Ruge-Murcia, Moto Shintani, and Toshi Watanabe, as well as participants at the 4th Hitotsubashi Summer Institute (HSI 2019), the 16th International Conference on Computational and Financial Econometrics (CFE 2022), the 2023 Spring Meetings of the Japanese Economic Association, the 9th Annual Conference of the International Association for Applied Econometrics (IAAE 2023), the 8th International Workshop on Financial Markets and Nonlinear Dynamics, the 2024 North American Summer Meetings of the Econometric Society (NASM 2024), EcoSta 2024, and the 2024 European Summer Meetings of the Econometric Society (ESEM 2024) for helpful comments and discussions.

I gratefully acknowledge financial support from JSPS KAKENHI (Grant Nos. 24330060, 17H02542, 17H00985, and 20H00073) and the Hitotsubashi Institute for Advanced Study. All remaining errors are my own.

\noindent\textit{Conflict of interest:} The author declares no conflict of interest.

\footnotesize

\newpage

\pagestyle{plain}

\renewcommand{\baselinestretch}{1.5}

%\doublespacing

\setcounter{footnote}{0}

\normalsize

\newpage

\begin{center}
{\large \textbf{1. Introduction}}
\end{center}

Structural macroeconomic models are abstractions of macroeconomic reality. They are designed to explain a limited set of economically meaningful mechanisms and population moments rather than every aspect of the joint distribution of the observed data. To preserve economic interpretation and theoretical discipline, they typically represent macroeconomic uncertainty through a parsimonious, low-dimensional set of structural shocks. A tension arises when such a model is evaluated using either a full-information likelihood or model-implied sampling distributions of selected sample moments, since both require a complete probability law for the observables. If the model is intended to explain only particular aspects of macroeconomic reality, the resulting parameter estimates and model comparisons may nevertheless reflect discrepancies in features of the data that lie outside the model’s intended explanatory scope.

Stochastic singularity is a form of misspecification in which the model has deficient stochastic rank: when there are fewer structural shocks than observables, the model may specify a complete but degenerate probability law on a lower-dimensional support. Because the data-generating joint distribution is nondegenerate, such a model is misspecified and cannot provide a usable full-information likelihood for the observed data. Conventional implementations restore full rank by adding shocks or measurement errors, but the added disturbances need not be compatible with the structural model's economic content.

The minimal econometric interpretation (MEI) introduced by \citet{geweke10} addresses this problem by abandoning the requirement that the structural model be a complete econometric model. It takes selected population moments, rather than observed data or sample moments, as the objects that the macroeconomic model is intended to explain. The structural model and its parameter prior induce prior distributions for these unobservable population moments, while a separate empirical reference model provides their posterior distributions conditional on the observed data. The structural model is assigned no direct probabilistic implications for the data or their sample moments and therefore need not specify their joint probability distribution. In this precise sense, MEI treats the structural macroeconomic model as an incomplete econometric model and permits model assessment without adding disturbances solely to remove stochastic singularity.

In Bayesian terms, MEI can be understood as a prior-predictive assessment conducted in the space of selected population moments. It evaluates the compatibility between the prior distributions of selected population moments implied by the structural model and their posterior distributions obtained from the empirical reference model, but it does not update the prior distribution of the structural parameters. This paper develops distribution-matching posterior inference (DMPI) to extend this comparison to posterior updating. DMPI converts distributional compatibility into a normalized score that, when combined with the structural-parameter prior, defines a generalized posterior in the sense of \citet{bissiri16}. It thereby turns MEI’s diagnostic comparison into posterior inference without requiring the structural model to specify a likelihood for the observables.

DMPI operationalizes this comparison using an exact
Dirichlet--multinomial (DM), or P\'{o}lya, marginal likelihood (the DM score)
for each selected population moment. Each DM score evaluates
compatibility between the discretized empirical and model-implied
distributions; their product is the DMPI score.
Although the moments are scored separately, their comparisons are linked
through the common structural-parameter ensemble: each parameter vector
generates the entire set of model-implied population moments.

The ensemble size \(M\) is a central inferential design parameter that controls the degree of structural discipline in DMPI. Under the $M$-fold product of the structural-parameter prior, increasing $M$ more tightly anchors the ensemble distribution to the distribution induced by that prior, as formalized by Sanov's large-deviation principle in Section~2.5. A larger \(M\) gives the model-implied counts greater weight in the DM construction. Each value of \(M\) defines a different generalized posterior. The normalizing constant of this generalized posterior, which we call the joint generalized evidence, serves two related purposes. At a common value of \(M\), it provides a basis for comparing different structural models. Within a given structural model, its profile over \(M\) indicates how empirical support changes as structural discipline increases.\footnote{The joint generalized evidence is analogous to a marginal likelihood, but it is conditional on the empirical population-moment target rather than being a marginal density of the raw data.}
 
Posterior inference and joint generalized evidence are computed using annealed
sequential Monte Carlo (SMC), which provides both posterior draws and
normalizing-constant estimates. This particle-based method is well suited to
high-dimensional, non-Gaussian, and multimodal DMPI posteriors; Section~2.6
provides the implementation details.

In sum, DMPI extends MEI from prior-predictive assessment to generalized
posterior inference and model comparison, using $M$-indexed structural
discipline to evaluate incomplete structural models without adding shocks or
measurement errors solely to obtain full rank.

Monte Carlo experiments based on a single-equation New Keynesian Phillips
Curve examine a stochastically singular reduced-rank (RR) model that omits one
shock present in the data-generating process. Its generalized-evidence profile
is hump-shaped: increasing $M$ initially raises the evidence, but further
increases cause the omitted-shock restriction to dominate.\footnote{The
correctly specified full-rank (FR) benchmark reported in Online Appendix~D has
a joint generalized-evidence profile that increases monotonically over the
examined grid of $M$.} A classical minimum-distance benchmark treats five
sample moments from each dataset as a point target and indicates that no
parameter vector can reproduce them jointly. Together, the results
distinguish exact validity from empirical usefulness: although the RR model is
misspecified as an exact representation, its theoretical restrictions remain
useful when imposed with moderate structural discipline.

The U.S. application compares one FR and three RR versions of
a three-equation New Keynesian DSGE model using posterior distributions of 21
selected population moments obtained from a Bayesian VAR. At common values of
$M$, the specification omitting the government-spending innovation has lower
generalized evidence than the FR model, whereas those omitting either
the technology or monetary-policy innovation have higher evidence. Thus,
adding a structural innovation can raise or lower the generalized evidence.
The set of included innovations therefore materially affects model comparison
in this application.

\vspace{3mm}
\noindent
\textbf{Related Literature:} This paper builds most directly on the MEI of
\citet{geweke10}.\footnote{MEI has been surveyed in the context of Bayesian
analysis of misspecified DSGE models by \citet{canova07},
\citet{dejongdave11}, \citet{delnegroschrfheide11}, and
\citet{fernandezvillaverde2016}.} Applications include
\citet{nasonrogers06} and \citet{kanonason14}; \citet{lmm22} use related
distributional comparisons to construct informative mixture priors. DMPI
extends this prior-predictive framework to generalized posterior inference on
structural parameters.

DMPI also connects the generalized Bayesian framework of \citet{bissiri16}
with composite likelihood methods \citep{varin11}: exact moment-specific
DM scores form its composite updating score.

The closest macroeconometric comparison is the DSGE-VAR framework of
\citet{delnegroschrfheide04}, which uses DSGE-implied population moment
matrices to construct a prior over the full parameterization of an otherwise
unrestricted VAR. DMPI instead applies structural discipline directly to
selected population-moment distributions. It can therefore evaluate
reduced-rank models without adding shocks or measurement errors solely to
obtain full rank.\footnote{\citet[footnote 8]{delnegroschrfheide04} discuss
the full-rank requirement and the addition of shocks or measurement errors
under stochastic singularity.} It also requires only that the selected moments
be computable, allowing nonlinear as well as linearized structural models.

Other points of comparison include classical minimum-distance estimation,
indirect inference
\citep{smith,gourieroux1993,gallanttauchen1996,dridietal2007}, Bayesian
indirect inference and Bayesian methods of moments
\citep{gallant_mculloch,gallant_giacomini_ragusa}, approximate Bayesian
computation \citep{marjoram03,forneron_ng18}, and limited-information
likelihood and quasi-Bayesian methods
\citep{kim02,ch03,schennach,is18}. These approaches typically update
structural parameters by comparing observed sample
statistics or moment conditions with model-implied or simulated counterparts.
In classical minimum-distance estimation, for example, sample moments are
point targets, with estimated sampling uncertainty entering through the
weighting matrix. DMPI instead carries sampling uncertainty through the
empirical posterior distributions of the selected population moments. Its
generalized-evidence profile over $M$ also compares prespecified degrees of
structural discipline, allowing structural restrictions to be imposed at an
empirically supported strength
rather than treated as simply accepted or rejected.

The remainder of the paper is organized as follows. Section 2 defines the DMPI generalized posterior, the exact DM (P\'{o}lya) marginal-likelihood score, the role of \(M\), and its implementation using annealed sequential Monte Carlo. Section 3 reports the New Keynesian Phillips Curve experiments and the omitted-shock diagnostic. Section 4 applies DMPI to one full-rank and three reduced-rank New Keynesian DSGE specifications using postwar U.S. data. Section 5 concludes. 

 \vspace{1mm}
\begin{center}
{\large \textbf{2. Distribution-Matching Posterior Inference (DMPI)}}
\end{center}

\vspace{1mm}
\noindent
\textit{2.1. Model roles and the inferential ensemble}

\vspace{1mm}

Under the MEI, the researcher selects $I$ population moments that the
structural model is intended to explain. These may be means, variances,
covariances, autocovariances, impulse-response ordinates, or other population
functionals.

Let $\mathbf y$ denote the observed data, and let $E$ be an empirical reference
model with parameter vector $\theta_E\in\Theta_E$ and posterior density
$p(\theta_E\mid\mathbf y,E)$. For each selected moment $i$,
$g_{E,i}(\theta_E)$ gives its value under $E$. Its posterior distribution
is obtained by applying $g_{E,i}$ to the empirical-model posterior and is
characterized by
\begin{equation}
 F_{E,i}(x\mid\mathbf y,E)
 =\Pr\{g_{E,i}(\theta_E)\leq x\mid\mathbf y,E\}
 =\int \mathbf 1\{g_{E,i}(\theta_E)\leq x\}
       p(\theta_E\mid\mathbf y,E)d\theta_E,
 \quad x\in\mathbb R.
 \label{eq:empirical_moment_posterior}
\end{equation}

Let $A$ denote a structural model with parameter vector $\theta_A\in
\Theta_A$ and proper prior density $\pi(\theta_A\mid A)$. Under $A$, the same
selected moment has model-implied value $g_{A,i}(\theta_A)$ for
$i=1,\ldots,I$. Both $g_{E,i}$ and $g_{A,i}$ may be nonlinear in their
respective parameters. Some parameter values do not produce a unique stable solution or finite values
of every declared population moment.  Let
\begin{equation*}
 \mathcal A_A=\{\theta_A:\text{$A$ has the required solution and every
 selected $g_{A,i}(\theta_A)$ is finite and well defined}\}.
\end{equation*}
The structural-parameter prior used by DMPI is defined on the admissible set as
\begin{equation}
 \pi_A(\theta_A)
 =\pi(\theta_A\mid A,\theta_A\in\mathcal A_A).
 \label{eq:conditional_prior}
\end{equation}
Parameter values outside $\mathcal A_A$ therefore have prior density zero.

The structural-parameter prior $\pi_A$ and the mappings $g_{A,i}$ induce a prior
distribution for each selected population moment, whereas
equation~(\ref{eq:empirical_moment_posterior}) gives the corresponding
posterior distribution under $E$ conditional on $\mathbf y$.  The role of
$E$ is to characterize empirical uncertainty about the population moment.

For a fixed positive integer $M$, let
\begin{equation*}
  \Theta_{A,M} =(\theta_A^1,\ldots,\theta_A^M), \quad \quad
  Q_M(C) =\frac{1}{M}\sum_{j=1}^M
  \mathbf 1\{\theta_A^j\in C\},\qquad C\subseteq\Theta_A.
\end{equation*}
Here, $\Theta_{A,M}$ is an ensemble of structural parameter vectors, and
$Q_M(C)$ is the fraction of ensemble members that lie in the region $C$.
Thus, $Q_M$ is the equally weighted distribution associated with the ensemble.
Applying $g_{A,i}$ to each member similarly gives the equally weighted
distribution of the model-implied values
$\{g_{A,i}(\theta_A^j)\}_{j=1}^M$ for the $i$th population moment.

The inferential state is the entire ordered ensemble $\Theta_{A,M}$, while
$Q_M$ is its permutation-invariant distributional representation.  The DMPI
construction treats the $M$ ensemble members symmetrically, so relabeling
them leaves the generalized posterior unchanged.  The posterior is therefore
exchangeable.
Consequently, all ensemble members have the same marginal posterior
distribution.  We define this common marginal---equivalently, the posterior
distribution of a uniformly selected ensemble member---as the DMPI posterior
for the structural-model parameter vector $\theta_A$. Online Appendix~B
verifies the exchangeability and common-marginal claims.

\vspace{1mm}
\noindent
\textit{2.2. A common discretization of empirical and model-implied population moment distributions}

\vspace{1mm}
Draw $N_{\mathrm{MC}}$ values of $\theta_E$ from
$p(\theta_E\mid\mathbf y,E)$ and transform each draw through
$g_{E,i}$. These draws approximate the empirical moment distribution. DMPI
maps that distribution into an effective count vector with total
$N_{\mathrm{eff}}$, specified separately from $N_{\mathrm{MC}}$.
$N_{\mathrm{MC}}$ controls the Monte Carlo approximation of the empirical
distribution, whereas $N_{\mathrm{eff}}$ sets the information scale entering
the structural update.
For each population moment $i$, let
$\mathcal B_i=\{B_{1,i},\ldots,B_{K,i}\}$ be a mutually exclusive and
exhaustive partition of the real line. The partition is constructed solely
from the posterior distribution of $g_{E,i}(\theta_E)$ under the empirical
model, fixed before structural-model comparison, and shared by all models.
The first and last cells are tail cells, so every finite model-implied value
is scored. Given posterior draws
$\{\theta_E^{(r)}\}_{r=1}^{N_{\mathrm{MC}}}$ from
$p(\theta_E\mid\mathbf y,E)$, define
\[
 \widehat p_{k,i}
 =
 \frac{1}{N_{\mathrm{MC}}}
 \sum_{r=1}^{N_{\mathrm{MC}}}
 \mathbf 1\!\left\{
 g_{E,i}(\theta_E^{(r)})\in B_{k,i}
 \right\},
 \qquad k=1,\ldots,K.
\]
Collect these cell probabilities in
$\widehat{\mathbf p}_i=(\widehat p_{1,i},\ldots,\widehat p_{K,i})'$.
The resulting probabilities are converted into the empirical count vector
\begin{equation}
 \mathbf n_i
 =
 \operatorname{LR}_{N_{\mathrm{eff}}}(\widehat{\mathbf p}_i)
 =
 (n_{1,i},\ldots,n_{K,i})',
 \qquad
 n_{k,i}\in\mathbb N_0,\qquad
 \sum_{k=1}^K n_{k,i}=N_{\mathrm{eff}},
 \label{multinomial}
\end{equation}
where $\operatorname{LR}_{N_{\mathrm{eff}}}$ denotes the largest-remainder
conversion.\footnote{The elements of
$N_{\mathrm{eff}}\widehat{\mathbf p}_i$ are first rounded down, and the
remaining units are assigned to the cells with the largest fractional
remainders. Hence $n_{k,i}/N_{\mathrm{eff}}\approx\widehat p_{k,i}$, while
the integer counts sum exactly to $N_{\mathrm{eff}}$.}

For the structural-parameter ensemble $\Theta_{A,M}$, define the corresponding
model-implied cell counts
\begin{equation}
 c_{k,i}(\Theta_{A,M})
 =\sum_{j=1}^M
   \mathbf 1\{g_{A,i}(\theta_A^j)\in B_{k,i}\},
 \qquad
 \sum_{k=1}^Kc_{k,i}(\Theta_{A,M})=M.
 \label{eq:structural_counts}
\end{equation}
The same ensemble member $\theta_A^j$ generates all $I$ population moments.
This shared-column construction is essential: it preserves the cross-equation
and cross-moment restrictions implied by $A$, even though the DMPI score below
is constructed from separate moment-specific DM scores.

\vspace{1mm}
\noindent
\textit{2.3. Two-stage updating and the exact DM score}

\vspace{1mm}
\enlargethispage{\baselineskip}
For each population moment $i$, introduce the latent cell-probability vector
$\mathbf p_i=(p_{1,i},\ldots,p_{K,i})'$ with the symmetric Dirichlet base
distribution
\begin{equation}
 \mathbf p_i\sim\mathrm{Dir}(\delta\mathbf 1_K),
 \qquad \delta>0.
 \label{eq:dirichlet_base}
\end{equation}
Here, $\mathbf p_i$ is a latent probability vector in the two-stage
conjugate construction.  It is distinct from the empirical frequency vector
$\widehat{\mathbf p}_i$ defined in Section~2.2, which has already been
converted into the fixed count vector $\mathbf n_i$.
Let
$\mathbf c_i(\Theta_{A,M})
=(c_{1,i}(\Theta_{A,M}),\ldots,c_{K,i}(\Theta_{A,M}))'$.
The model-implied counts in equation~(\ref{eq:structural_counts}) provide the first conjugate update:
\begin{equation}
 \mathbf p_i\mid\Theta_{A,M},A
 \sim
 \mathrm{Dir}\{\boldsymbol\alpha_i(\Theta_{A,M})\},
 \qquad
 \boldsymbol\alpha_i(\Theta_{A,M})
 =
 \delta\mathbf 1_K+\mathbf c_i(\Theta_{A,M}).
 \label{dirichlet}
\end{equation}
Thus,
$\alpha_{k,i}=\delta+c_{k,i}$ and
$\alpha_{0,i}=\sum_k\alpha_{k,i}=M+K\delta$.
The empirical counts in equation~(\ref{multinomial}) provide the second update:
\begin{equation}
 \mathbf p_i\mid\Theta_{A,M},\mathbf n_i,A,E
 \sim
 \mathrm{Dir}\{\boldsymbol\alpha_i(\Theta_{A,M})+\mathbf n_i\}.
 \label{eq:dirichlet_data_update}
\end{equation}
The three sources of mass have distinct roles: $K\delta$ provides base
regularization, $M$ is the structural concentration, and
$N_{\mathrm{eff}}$ is the empirical information scale.

Integrating $\mathbf p_i$ out of the multinomial sampling model gives the exact
Dirichlet--multinomial (DM), or P\'{o}lya, probability mass function
\citep{lindley64}, which defines the moment-specific DM score:
\begin{align}
 L_{i,M}(\Theta_{A,M})
 &\equiv
 p_{\mathrm{DM}}(\mathbf n_i\mid\Theta_{A,M}) \notag\\
 &=
 \frac{\Gamma(N_{\mathrm{eff}}+1)\Gamma(M+K\delta)}
      {\Gamma(N_{\mathrm{eff}}+M+K\delta)}
 \prod_{k=1}^K
 \frac{\Gamma\{n_{k,i}+c_{k,i}(\Theta_{A,M})+\delta\}}
      {\Gamma(n_{k,i}+1)
       \Gamma\{c_{k,i}(\Theta_{A,M})+\delta\}}.
 \label{polya}
\end{align}
A conjugate derivation is provided in Online Appendix A. The additive
smoothing term $\delta>0$ gives every cell positive base mass, including
cells with zero model-implied count. A positive empirical count in such a
cell therefore does not force the moment-specific DM score to zero, even when
the empirical and model-implied moment distributions do not overlap. The
incompatible moment remains in the score and its mismatch is penalized, but it
cannot act as a zero-score veto on posterior inference and model comparison.
Equation~(\ref{polya}) is evaluated in log space using standard log-gamma
routines.

\vspace{1mm}
\noindent
\textit{2.4. The generalized posterior and joint generalized evidence}

\vspace{1mm}
The DMPI score combines the moment-specific DM scores
across the selected population moments:
\begin{equation}
 L_M(\Theta_{A,M})
 =\prod_{i=1}^I L_{i,M}(\Theta_{A,M}).
 \label{eq:dmpi_core}
\end{equation}
Let $\mathcal T_E(\mathbf y)=\{(\mathcal B_i,\mathbf n_i)\}_{i=1}^I$ denote the
fixed empirical target constructed from
$p(\theta_E\mid\mathbf y,E)$ as described in Section~2.2, for the selected
population moments and fixed values of $N_{\mathrm{eff}}$ and $K$. The
smoothing parameter $\delta$ is also fixed before structural inference. The
fixed target argument is suppressed in $L_{i,M}$, $L_M$, and $Z_{A,M}$ below;
for example, $L_M(\Theta_{A,M}) \equiv
L_M\bigl(\Theta_{A,M};\mathcal T_E(\mathbf y)\bigr)$.
Thus, the observed data enter the structural update through the
empirical target, while the empirical-model posterior remains fixed during
the structural update.

The product structure in equation~(\ref{eq:dmpi_core}) implements the scope
restriction of the MEI.  The structural model is evaluated
through the marginal distribution of each selected population moment that it
is intended to explain.  It is not required to reproduce the joint sampling
distribution, including the sampling covariances, of the corresponding sample
moments.  In statistical terms, $L_M$ has
the form of a composite marginal score \citep{varin11}, although its
components are DM scores rather than likelihood
contributions from a fully specified sampling model.

Let $\Pi_{A,M}^{0}$ denote the product prior measure on the
structural-parameter ensemble, and let $\pi_{A,M}^{0}$ denote its density with
respect to $d\Theta_{A,M}$.  Independence of the ensemble members under the
structural prior implies
\begin{equation}
 \Pi_{A,M}^{0}(d\Theta_{A,M})
 =\pi_{A,M}^{0}(\Theta_{A,M})\,d\Theta_{A,M},
 \qquad
 \pi_{A,M}^{0}(\Theta_{A,M})
 =\prod_{j=1}^M\pi_A(\theta_A^j).
 \label{direct_prior}
\end{equation}
Thus, $\pi_{A,M}^{0}$ is the $M$-fold product density induced by the
structural prior $\pi_A$ defined in
equation~(\ref{eq:conditional_prior}).
Combining the score in equation~(\ref{eq:dmpi_core}) with the product prior
gives the generalized posterior measure $\Pi_{A,M}$, whose density is
\begin{equation*}
 \begin{aligned}
 \pi_{A,M}(\Theta_{A,M}\mid\mathbf y,A,E)
 &\equiv
 \pi_{A,M}(\Theta_{A,M}\mid\mathcal T_E(\mathbf y),A)\\
 &=
 \frac{
  L_M(\Theta_{A,M})
  \pi_{A,M}^{0}(\Theta_{A,M})
 }{
  Z_{A,M}
 }.
 \end{aligned}
\end{equation*}
The normalizing constant is
\begin{equation*}
 \begin{aligned}
 Z_{A,M}
 &=\int
   L_M(\Theta_{A,M})
   \pi_{A,M}^{0}(\Theta_{A,M})\,d\Theta_{A,M}\\
 &=\mathbb E_{\Theta_{A,M}\sim\Pi_{A,M}^{0}}
   \left[
    \prod_{i=1}^I
    L_{i,M}(\Theta_{A,M})
   \right].
 \end{aligned}
\end{equation*}
We call $Z_{A,M}$ the \textit{joint generalized evidence} conditional on structural admissibility. It is the normalizing constant of the generalized posterior and plays a role analogous to the marginal likelihood in ordinary Bayesian inference, although it is not a marginal density of the raw data $\mathbf y$. The term generalized evidence is used here in this specific sense. Using exact DM probabilities in $L_M$ fixes the numerical scale of $Z_{A,M}$ under the declared count construction. It is joint because the product of the population moment-specific scores is integrated once over the shared structural-parameter ensemble.

For two structural models $A_1$ and $A_2$, model comparison at a common value
of $M$ uses
\begin{equation*}
 \Delta_M(A_1,A_2)
 =\log Z_{A_1,M}-\log Z_{A_2,M}.
\end{equation*}
A positive value favors $A_1$. Because equation~(\ref{eq:dmpi_core}) defines
a declared population-moment score rather than a likelihood for $\mathbf y$,
$\Delta_M$ is not an ordinary log Bayes factor. Instead,
$\exp\{\Delta_M(A_1,A_2)\}=Z_{A_1,M}/Z_{A_2,M}$ is a general Bayes factor in
the sense of \citet{bissiri16}.

\vspace{1mm}
\noindent
\textit{2.5. The product prior, the discipline index \(M\), and boundary cases}

\vspace{1mm}
The product prior in equation~(\ref{direct_prior}) induces a prior distribution
over $Q_M$, the distributional representation of the structural-parameter
ensemble.  To characterize this implication, consider any finite measurable
partition $\{C_1,\ldots,C_L\}$ of the parameter space and define
\begin{equation*}
 \bar\pi_l=\int_{C_l}\pi_A(\theta_A)\,d\theta_A,
 \qquad
 r_l=M Q_M(C_l)
 =\sum_{j=1}^M\mathbf 1\{\theta_A^j\in C_l\},
 \qquad l=1,\ldots,L.
\end{equation*}
Let $\bar{\boldsymbol\pi}=(\bar\pi_1,\ldots,\bar\pi_L)'$ and
$\mathbf r=(r_1,\ldots,r_L)'$.  Under the product prior, $\mathbf r$ has the
exact multinomial probability mass function
\begin{equation}
 \Pr_{\Pi_{A,M}^{0}}(\mathbf r)
 =\frac{M!}{\prod_{l=1}^Lr_l!}
  \prod_{l=1}^L\bar\pi_l^{r_l},
 \qquad \sum_{l=1}^Lr_l=M.
 \label{eq:sanov_finite}
\end{equation}
The method of types \citep[Theorem~11.1.4]{coverthomas06} gives
\begin{equation}
 \log \Pr_{\Pi_{A,M}^{0}}(\mathbf r)
 =-M D_{\mathrm{KL}}(\mathbf r/M\Vert\bar{\boldsymbol\pi})
 +O(\log M).
 \label{eq:sanov_anchor}
\end{equation}
Here
$D_{\mathrm{KL}}(\mathbf w\Vert\bar{\boldsymbol\pi})
=\sum_{l=1}^{L}w_l\log(w_l/\bar\pi_l)$ denotes the Kullback--Leibler
divergence, with $0\log 0=0$.
Let $\mathbf w=\mathbf r/M$.  Equation~(\ref{eq:sanov_anchor}) shows that, for
any fixed partition, a cell-frequency vector $\mathbf w$ that differs from the
prior probabilities $\bar{\boldsymbol\pi}$ receives prior probability of order
$\exp\{-M D_{\mathrm{KL}}(\mathbf w\Vert\bar{\boldsymbol\pi})\}$.  Thus, such
deviations become exponentially less likely as $M$ increases.  This is the
finite-partition form of Sanov's large-deviation principle \citep{sanov57};
Online Appendix B gives the corresponding derivation. The multinomial
coefficient in equation~(\ref{eq:sanov_finite}) counts the assignments of the
$M$ labeled ensemble members to partition cells that yield the same
cell-frequency vector. Thus, the product prior itself induces this
distribution-level concentration.

This argument also clarifies the role of $M$. For a given profile of ensemble
cell frequencies, increasing $M$ raises the total Dirichlet concentration
parameter $M+K\delta$ in equation~(\ref{dirichlet}) and reduces the conditional
variance of the latent cell probabilities around their smoothed means. At the
same time, under the product prior, ensemble distributions that remain
separated from the distribution induced by $\pi_A$ become exponentially less
probable as $M$ increases. In these two respects, $M$ controls the degree of
structural discipline imposed by DMPI. Thus, for fixed $N_{\mathrm{eff}}$,
$K$, $\delta$, and $\{\mathcal B_i\}_{i=1}^I$, each value of $M$ defines a
different generalized posterior and joint generalized evidence.

Two boundary cases clarify the role of $M$. First, let $M=1$ and denote by
$h_i$ the index of the cell containing the single model-implied value
$g_{A,i}(\theta_A^1)$. Then $c_{h_i,i}=1$ and $c_{k,i}=0$ for
$k\neq h_i$. Substitution into equation~(\ref{polya}) gives
\[
 L_{i,1}(\theta_A^1)\propto n_{h_i,i}+\delta,
\]
where the omitted factor is constant with respect to $\theta_A^1$. Thus, the
$M=1$ score favors parameter values whose model-implied population moment falls in a cell
receiving greater empirical mass, while $\delta$ prevents cells with zero
empirical counts from receiving zero score. If $N_{\mathrm{eff}}=1$ and $e_i$
denotes the index of the cell in which the empirical count vector places its
single unit, then
\[
 L_{i,1}(\theta_A^1)
 =\frac{\mathbf 1\{h_i=e_i\}+\delta}{1+K\delta}
 \longrightarrow \mathbf 1\{h_i=e_i\}
 \qquad\text{as }\delta\downarrow0.
\]
Thus, the limiting score is a discretized analogue of a hard-tolerance ABC
kernel: it assigns unit weight when the model-implied and empirical values
occupy the same cell and zero weight otherwise.

Second, let $M\rightarrow\infty$ with the empirical target and $\delta$
fixed. Under the product prior, the strong law of large numbers gives, for
each $k$ and $i$,
\begin{equation*}
 \frac{c_{k,i}(\Theta_{A,M})}{M}
 \xrightarrow{\mathrm{a.s.}}
 p_{A,k,i}
 \equiv
 \Pr_{\theta_A\sim\pi_A}
 \{g_{A,i}(\theta_A)\in B_{k,i}\}.
\end{equation*}
Let $\mathbf p_{A,i}=(p_{A,1,i},\ldots,p_{A,K,i})'$. This vector should be
distinguished from the latent cell-probability vector $\mathbf p_i$ introduced
in equation~(\ref{eq:dirichlet_base}) and structurally updated in
equations~(\ref{dirichlet}) and (\ref{eq:dirichlet_data_update}). Whereas $\mathbf p_i$ is random in the finite-$M$
Dirichlet--multinomial construction, $\mathbf p_{A,i}$ is fixed once $A$,
$\pi_A$, and the partition $\mathcal B_i$ are specified. Their relationship is
\[
 \mathbb E(\mathbf p_i\mid\Theta_{A,M},A)
 =
 \frac{\mathbf c_i(\Theta_{A,M})+\delta\mathbf 1_K}
      {M+K\delta}
 \xrightarrow{\mathrm{a.s.}}
 \mathbf p_{A,i}.
\]
Moreover, because the total Dirichlet concentration $M+K\delta$ diverges, the
conditional distribution of $\mathbf p_i$ concentrates at
$\mathbf p_{A,i}$.

As shown in Online Appendix B, provided that $p_{A,k,i}>0$ whenever
$n_{k,i}>0$, the DM score satisfies
\begin{equation*}
 L_{i,M}(\Theta_{A,M})
 \xrightarrow{\mathrm{a.s.}}
 \frac{N_{\mathrm{eff}}!}{\prod_{k=1}^K n_{k,i}!}
 \prod_{k=1}^K p_{A,k,i}^{n_{k,i}}.
\end{equation*}
Because $0\leq L_{i,M}\leq1$ and the limiting DM scores are deterministic,
bounded convergence gives
\begin{equation*}
 \lim_{M\rightarrow\infty}Z_{A,M}
 =
 \prod_{i=1}^I
 \left\{
 \frac{N_{\mathrm{eff}}!}{\prod_{k=1}^K n_{k,i}!}
 \prod_{k=1}^Kp_{A,k,i}^{n_{k,i}}
 \right\}.
\end{equation*}

With the empirical target fixed, posterior reweighting does not overturn the
product-prior concentration: the score-relevant cell frequencies $c_{k,i}/M$
still converge to their prior-predictive probabilities $p_{A,k,i}$; Online
Appendix B gives the argument. The limiting evidence nevertheless remains a
nontrivial prior-predictive compatibility score between the structural and
empirical population-moment distributions. It is therefore the discretized
counterpart of the model-evaluation comparison underlying MEI. DMPI thus
provides an $M$-indexed bridge between finite-$M$ posterior distribution
matching and the MEI prior-predictive endpoint.

The interpretation of the $M$ profile depends on the inferential question.
For model comparison, $Z_{A,M}$ should be compared across structural models at
a common value of $M$, so that the degree of structural discipline is held
fixed. Within a given model, variation in $Z_{A,M}$ across $M$ serves as a
specification diagnostic, showing how empirical compatibility changes as
structural discipline is varied. Parameter summaries at any particular $M$
are conditional on that degree of discipline.

Let $\mathcal M$ denote a finite set of ensemble sizes fixed before examining
the evidence results. For a given structural model $A$, define the
evidence-supported degree of structural discipline as
\begin{equation*}
 M_A^\star
 =
 \arg\max_{M\in\mathcal M} Z_{A,M}.
\end{equation*}
Thus, $M$ can be learned from the declared empirical target in an
empirical-Bayes sense.

\vspace{1mm}
\noindent
\textit{2.6. Annealed SMC for generalized posterior inference and evidence}

\vspace{1mm}
For each structural model $A$ and ensemble size $M$, DMPI uses an annealed
SMC sampler \citep{delmoral06,hs15} with tempered targets
\begin{equation*}
 \Pi_{A,M,t}(d\Theta_{A,M})
 \propto
 L_M(\Theta_{A,M})^{\beta_t}
 \Pi_{A,M}^{0}(d\Theta_{A,M}),
 \qquad
 0=\beta_0<\beta_1<\cdots<\beta_T=1.
\end{equation*}
Thus, $\Pi_{A,M,0}=\Pi_{A,M}^{0}$ and
$\Pi_{A,M,T}=\Pi_{A,M}$. The annealing sequence
$\{\beta_t\}_{t=0}^T$ is a computational device and does not alter the final
generalized posterior.

Let $P$ denote the number of SMC particles, and let
$\{\Theta_{t-1}^{p}\}_{p=1}^P$ denote the equally weighted particles from the
previous stage, with the subscripts $A$ and $M$ suppressed. The incremental
weights are
\begin{equation*}
 \widetilde w_t^p
 =L_M(\Theta_{t-1}^{p})^{\beta_t-\beta_{t-1}}.
\end{equation*}
At each stage, the algorithm records the effective sample size of these
weights, evaluates the logarithm of their mean by the log-sum-exp identity and
adds it to the log normalizing-constant estimate, performs systematic
resampling, and applies mutation kernels invariant to $\Pi_{A,M,t}$. The
resulting evidence estimator is
\begin{equation}
 \log\widehat Z_{A,M}
 =\sum_{t=1}^T
  \log\left\{\frac{1}{P}\sum_{p=1}^P\widetilde w_t^p\right\}.
 \label{eq:smc_logz}
\end{equation}
The SMC counterpart of $M_A^\star$ is
$\widehat M_A=\arg\max_{M\in\mathcal M}\widehat Z_{A,M}$.
Online Appendix C derives the normalizing-constant identity in
equation~(\ref{eq:smc_logz}) and gives the Metropolis--Hastings acceptance
probability for the mutation kernels described below.

Mutation updates the structural-parameter ensemble one column at a time. When
column $j$ is replaced by a proposal $\theta_A'$, all other columns are held
fixed. Because the ensemble prior has the product form in
equation~(\ref{direct_prior}), its contribution to the log
Metropolis--Hastings ratio reduces to $ \log\pi_A(\theta_A')-\log\pi_A(\theta_A^j)$.
Thus, the change in the log prior involves only the proposed column and does
not grow with the ensemble size $M$. The DMPI score $L_M$ is updated
to reflect the changes in the model-implied cell counts generated by that
column.\footnote{The SMC sampler combines local random-walk proposals,
prior-independence proposals, and a global Student-$t$ proposal fitted in
transformed parameter space. The global proposal is fixed within each
annealing stage. Metropolis--Hastings ratios include the proposal-density and
required Jacobian terms. A preliminary calculation determines the temperature
schedule, which is fixed across independent runs.}

Table~\ref{tab:information_accounting} distinguishes target-defining from
computational quantities; Online Appendix E reports the numerical diagnostics
for the SMC estimates.

\begin{table}[H]
\centering
\footnotesize
\setlength{\tabcolsep}{3pt}
\renewcommand{\arraystretch}{0.92}
\begin{tabular}{@{}
 >{\raggedright\arraybackslash}p{0.25\textwidth}
 >{\raggedright\arraybackslash}p{0.45\textwidth}
 >{\raggedright\arraybackslash}p{0.16\textwidth}
 @{}}
\hline
Quantity or setting & Role & Class \\
\hline
$M$
& Ensemble size and structural discipline
& Target \\

$\{\mathbf n_i,\mathcal B_i\}_{i=1}^I$,
$N_{\mathrm{eff}}$, $K$, $\delta$
& Empirical target, information scale, discretization, and smoothing
& Target \\

$\pi_A$, $\mathcal A_A$
& Structural prior and admissible support
& Target \\

$N_{\mathrm{MC}}$
& Accuracy of empirical-target construction
& Computation \\

$P$, $\{\beta_t\}_{t=0}^T$
& SMC accuracy and tempering path
& Computation \\
\hline
\end{tabular}
\captionsetup{font=small}
\caption{Information accounting in DMPI.}
\label{tab:information_accounting}
\end{table}

\vspace{1mm}
\begin{center}
{\large \textbf{3. Monte Carlo Experiments with a Stochastically Singular \\
New Keynesian Phillips Curve Model}}
\end{center}

\vspace{1mm}
This section studies DMPI in a simple environment in which stochastic
singularity is transparent. Data are
generated from the FR NKPC model with two
observables and two structural shocks. The main analysis uses an RR
version that omits the NKPC shock and retains only the output-gap
shock. Because it has fewer structural shocks than observables, this model is
stochastically singular.

The experiment examines how DMPI behaves when the omitted shock prevents the
structural model from jointly reproducing the empirical distributions of all
declared population moments. For comparison, Online Appendix~D reports the
corresponding benchmark experiment in which the structural model includes both
shocks and is correctly specified.

\vspace{1mm}
\noindent 
\textit{3.1. The data-generating process and the RR NKPC}

\vspace{1mm}
The data-generating process is
\begin{equation}
\Delta\pi_t
=\beta\mathbf E_t\Delta\pi_{t+1}+\kappa\phi_t+v_t,
\qquad
v_t\sim\mathrm{i.i.d.}\ \mathcal N(0,\sigma_v^2),
\label{NKPC}
\end{equation}
where $\Delta\pi_t$ is the change in inflation, $\phi_t$ is the output gap,
$\mathbf E_t$ conditions on time-$t$ information, and $v_t$ is the NKPC
shock. The subjective discount factor is $\beta$, and the slope of the NKPC is
\begin{equation*}
\kappa=\frac{(1-\mu_p)(1-\beta\mu_p)}{\mu_p},
\end{equation*}
where $\mu_p$ is the Calvo probability of price non-adjustment. The exogenous
output gap follows
\begin{equation}
\phi_t=\rho\phi_{t-1}+\epsilon_t,
\qquad
\epsilon_t\sim\mathrm{i.i.d.}\ \mathcal N(0,\sigma_\epsilon^2).
\label{output_gap}
\end{equation}

Let
\begin{equation*}
b(\beta,\mu_p,\rho)=\frac{\kappa}{1-\beta\rho}.
\end{equation*}
Under the unique fundamental solution, equations~(\ref{NKPC}) and
(\ref{output_gap}) imply
\begin{equation*}
\begin{bmatrix}
\Delta\pi_t \\
\phi_t
\end{bmatrix}
=
\begin{bmatrix}
0 & b\rho \\
0 & \rho
\end{bmatrix}
\begin{bmatrix}
\Delta\pi_{t-1} \\
\phi_{t-1}
\end{bmatrix}
+
\begin{bmatrix}
b & 1 \\
1 & 0
\end{bmatrix}
\begin{bmatrix}
\epsilon_t \\
v_t
\end{bmatrix}
\end{equation*}
The empirical reference model $E$ is the corresponding unrestricted VAR:
\begin{equation}
\begin{bmatrix}
\Delta\pi_t \\
\phi_t
\end{bmatrix}
=
\begin{bmatrix}
a_{11}  & a_{12} \\
a_{21} &  a_{22}
\end{bmatrix}
\begin{bmatrix}
\Delta \pi_{t-1} \\
\phi_{t-1}
\end{bmatrix}
+ 
\mathbf e_t,
\qquad
\mathbf e_t\sim\mathrm{i.i.d.}\ \mathcal N(\mathbf 0,\Sigma_e).
\label{unrestrictedVAR}
\end{equation}
For each draw of the empirical-model parameter vector, let
$\mathbf g_E(\theta_E)=(g_{E,1}(\theta_E),\ldots,g_{E,5}(\theta_E))'$ denote
the five selected population moments.

The structural model $A$ used for inference
omits $v_t$ and has parameter vector
$\theta_A=(\beta,\mu_p,\rho,\sigma_\epsilon^2)'$; define its model-implied
counterpart $\mathbf g_A(\theta_A)$ analogously. The two mappings are
\begin{align}
\mathbf g_E(\theta_E)
&=\bigl(a_{12},a_{22},[\Sigma_e]_{11},[\Sigma_e]_{12},[\Sigma_e]_{22}\bigr)',
\label{eq:nkpc_empirical_features}\\
\mathbf g_A(\theta_A)
&=\bigl(b\rho,\rho,b^2\sigma_\epsilon^2,
         b\sigma_\epsilon^2,\sigma_\epsilon^2\bigr)'.
\label{eq:nkpc_structural_features}
\end{align}
At the common parameter values, the FR DGP instead implies
$[\Sigma_e]_{11}=b^2\sigma_\epsilon^2+\sigma_v^2$. Thus, the omitted shock
creates a transparent discrepancy in the third target moment while leaving
the other four structural relationships unchanged. The reduced-form innovation covariance matrix implied by the structural model
used for inference has rank one. Nevertheless, the two autoregressive
coefficients and three covariance elements defining the five
population-moment mappings in equation~(\ref{eq:nkpc_structural_features})
remain finite and well defined.

\vspace{1mm}
\pagebreak[4]
\noindent
\textit{3.2. Monte Carlo design and implementation}

\vspace{1mm}
The FR DGP is calibrated at
$\beta_0=0.98$, $\mu_{p,0}=0.8$, $\rho_0=0.8$,
$\sigma_{\epsilon,0}^2=0.001$, and $\sigma_{v,0}^2=0.00025$. For each Monte
Carlo replication, we initialize the process at zero and simulate 30{,}000
observations of $(\Delta\pi_t,\phi_t)'$. We discard the first 28{,}000
observations as burn-in and retain the following $T=300$ observations as
$\mathbf y$. The structural prior for the
four parameters of the RR model is informative and centered at the
corresponding DGP values, as reported in Table~\ref{tab:prior_nkpc}. This
design isolates the effect of the omitted NKPC shock from prior
misspecification.

\begin{table}[H]
\renewcommand{\arraystretch}{0.7}
\centering
\captionsetup{font=footnotesize}  
\caption{Prior distributions for the RR NKPC model}
\label{tab:prior_nkpc}
\scriptsize

\begin{tabular}{cccc}
\hline
\hline
\textbf{Name} & \textbf{Density} & \textbf{Mean} & \textbf{SD} \\
\hline
$\beta$  & Beta          & 0.980 & 0.0010 \\
$\mu_p$   & Beta           & 0.800 & 0.0316 \\
$\rho$     & Beta           & 0.800 & 0.0316 \\
$\sigma_\epsilon^2$    & Truncated Normal & 0.001 & 0.0001 \\

\hline
\multicolumn{4}{p{0.4\linewidth}}{\scriptsize \textit{Note.} All truncated normal priors are truncated to positive support.} 
\end{tabular}
\end{table}

For each replication, we draw directly from the conjugate
normal--inverse-Wishart posterior of the unrestricted VAR in
equation~(\ref{unrestrictedVAR}). We obtain $N_{\mathrm{MC}}=50{,}000$
independent parameter draws and apply the five mappings in
equation~(\ref{eq:nkpc_empirical_features}) to every draw. The transformed draws 
approximate the five empirical population-moment distributions conditional on the simulated data.

Each distribution is discretized into $K=300$ cells. For $a_{12}$ and
$a_{22}$, the bounded interior cells cover $[0,2]$. The corresponding ranges
for $[\Sigma_e]_{11}$, $[\Sigma_e]_{12}$, and $[\Sigma_e]_{22}$ are
$[0,0.015]$, $[0,0.005]$, and $[0,0.005]$, respectively. In every partition,
the first and last cells extend to $-\infty$ and $+\infty$, so values outside
these interior ranges remain part of the target. The empirical cell
probabilities are converted deterministically into integer counts with
$N_{\mathrm{eff}}=N_{\mathrm{MC}}=50{,}000$. This preserves the simulated cell
counts exactly, avoiding additional coarsening of the empirical
population-moment distributions. The counts and partitions are then fixed
before structural inference.

We set the symmetric Dirichlet concentration to $\delta=1$ and consider
ensemble sizes $\mathcal M$ =\{1, 10, 50, 100, 200, 300\}. For each $M \in \mathcal M$, we compute the
generalized posterior and joint generalized evidence using the exact
DM score and the annealed SMC procedure described in
Section~2.6. The reported calculations use $P=512$ particles and three mutation sweeps at
each annealing stage. The temperature schedules are determined in preliminary
calculations and then held fixed. The reported results average estimates
based on $P=512$ particles across 30 independently simulated datasets.
Separate, prespecified runs with an independent seed at $P=512$ and with
$P=1024$ assess numerical stability.

\vspace{1mm}
\noindent 
\textit{3.3. Results for the RR model}

\vspace{1mm}
Table~\ref{tab:posterior_misspec} reports Monte Carlo averages of posterior
summaries for the structural parameters and the estimated log joint
generalized evidence for the RR model. Across all values of $M$, the
posterior means of $\beta$, $\mu_p$, $\rho$, and $\sigma_\epsilon^2$ remain
close to their DGP values. Posterior intervals do not narrow as $M$ increases:
the interval for $\beta$ remains nearly
unchanged, whereas those for $\mu_p$, $\rho$, and $\sigma_\epsilon^2$ widen
moderately. This is consistent with $M$ governing structural discipline rather
than empirical information.

Although the RR model cannot jointly reconcile the empirical
distribution of $[\Sigma_e]_{11}$ with the remaining population moments, this
incompatibility does not produce large systematic displacements of the
posterior means. Two features of the specification help explain this result.
Laplace smoothing with $\delta=1$ ensures that the mismatch in
$[\Sigma_e]_{11}$ does not force the DMPI score to zero for otherwise
admissible ensembles. The centered informative prior plays a complementary
role by discouraging large parameter shifts that would improve the fit to this moment
only at the cost of worsening the fit to the remaining population moments.
Together, these mechanisms allow the incompatibility to remain localized.

\begin{table}[H]
\centering
\captionsetup{font=footnotesize}  
\caption{Monte Carlo results for the RR NKPC model}
\label{tab:posterior_misspec}
\scriptsize
\setlength{\tabcolsep}{4pt}
\renewcommand{\arraystretch}{0.78}
\begin{tabular}{c|c|c|c|c|c}
\hline
$M$ & $\beta$ & $\mu_p$ & $\rho$ & $\sigma_\epsilon^2$ & $\log\widehat Z_{\mathrm{RR},M}$ \\
 & [0.980] & [0.800] & [0.800] & [0.001] & \\
\hline\hline
1 & 0.98000 & 0.80160 & 0.79718 & 0.000999 & -9128.71 \\
  & (0.97805, 0.98191) & (0.77418, 0.82826) & (0.74843, 0.84342) & (0.000888, 0.001119) & (0.81) \\
\hline
10 & 0.97999 & 0.80165 & 0.79726 & 0.001002 & -9089.51 \\
   & (0.97799, 0.98190) & (0.77030, 0.83147) & (0.74515, 0.84599) & (0.000879, 0.001134) & (8.17) \\
\hline
50 & 0.98000 & 0.80122 & 0.79654 & 0.001004 & -9027.58 \\
   & (0.97799, 0.98191) & (0.76464, 0.83535) & (0.73676, 0.85241) & (0.000862, 0.001156) & (41.17) \\
\hline
100 & 0.98000 & 0.80089 & 0.79639 & 0.001005 & -9066.29 \\
    & (0.97799, 0.98191) & (0.76148, 0.83671) & (0.73313, 0.85538) & (0.000855, 0.001166) & (82.15) \\
\hline
200 & 0.98000 & 0.80036 & 0.79638 & 0.001005 & -9295.66 \\
    & (0.97799, 0.98191) & (0.75756, 0.83761) & (0.73013, 0.85787) & (0.000850, 0.001174) & (162.68) \\
\hline
300 & 0.98000 & 0.79999 & 0.79638 & 0.001005 & -9616.25 \\
    & (0.97799, 0.98191) & (0.75498, 0.83794) & (0.72880, 0.85895) & (0.000847, 0.001178) & (241.84) \\
\hline
\end{tabular}
\vspace{1mm}
\captionsetup{font=scriptsize}
\caption*{\textit{Note.} Values in brackets are the DGP parameters. For each
structural parameter, the table reports the Monte Carlo average of the
posterior mean (top) and the Monte Carlo averages of the 2.5th
and 97.5th posterior percentiles (bottom). The final column reports the Monte
Carlo mean of the estimated log joint generalized evidence, with its Monte
Carlo standard deviation in parentheses. Results are based on 30 independently
simulated datasets.}
\end{table}

\begin{figure}[H]
    \centering
    \includegraphics[width=0.96\linewidth]{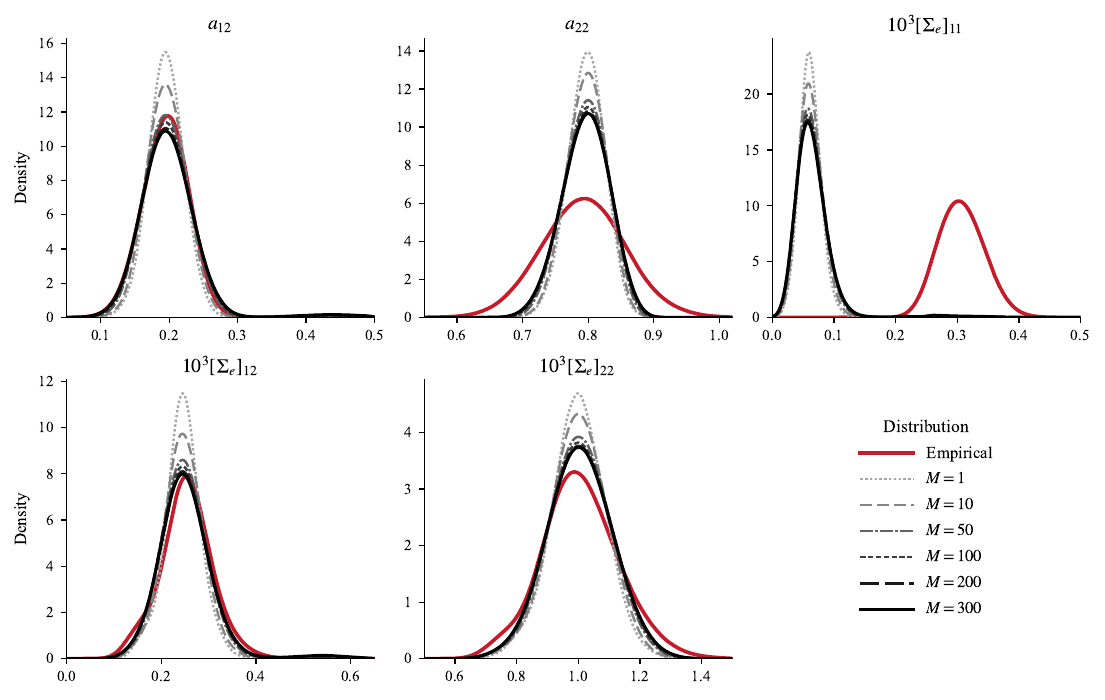}
    \captionsetup{font=footnotesize}
    \caption{Monte Carlo-averaged population-moment distributions for the
    misspecified RR NKPC model}
    \label{fig:mis_correct_prior_theory_empirical}
    \captionsetup{font=scriptsize}
    \caption*{\textit{Note.} The empirical curve pools the 50{,}000
    VAR-posterior moment draws from each of 30 simulated datasets. For each
    $M$, the model-implied curve pools the members of the final equally
    weighted SMC ensembles across the same datasets. All curves use a common
    smoothing bandwidth within each panel. The three covariance moments are
    multiplied by $10^3$.}
\end{figure}

\vspace{-5mm}
Figure~\ref{fig:mis_correct_prior_theory_empirical} compares the Monte
Carlo-averaged empirical and model-implied densities for the five selected
population moments. The dominant regions of the model-implied distributions
for $a_{12}$, $a_{22}$, $[\Sigma_e]_{12}$, and $[\Sigma_e]_{22}$ overlap
substantially with their empirical counterparts, although some differences in
dispersion remain. By contrast, the model-implied distribution of
$[\Sigma_e]_{11}$, the moment directly affected by the omitted NKPC shock,
is centered well below its empirical distribution. For $M\leq50$, the two
densities exhibit negligible overlap. For $M\geq100$, a small secondary
component moves toward the empirical high-density region, but the dominant
model-implied mass remains separated from it.

Because the same ensemble member generates all five moments, this adjustment
cannot occur in $[\Sigma_e]_{11}$ alone. The emerging right-hand component is
accompanied by corresponding right-tail components in $a_{12}$ and
$[\Sigma_e]_{12}$. The figure therefore illustrates the structural tradeoff:
improving the fit of the omitted-shock moment requires disturbing moments
that the RR model otherwise matches relatively well.

The final column of Table~\ref{tab:posterior_misspec} reports the Monte Carlo
mean of the log joint generalized evidence. The mean increases from
$-9128.71$ at $M=1$ to $-9027.58$ at $M=50$, and then declines to
$-9616.25$ at $M=300$. Among the values considered, $M=50$ yields the largest
evidence in 24 of the 30 simulated datasets; the remaining maxima occur at
$M=10$ or $M=100$. The decline is especially pronounced beyond $M=100$.

This hump-shaped $M$ profile is the central within-model specification
diagnostic in DMPI. Increasing $M$ raises the structural concentration and
strengthens the product-prior anchoring of the ensemble distribution. At
intermediate values of $M$, the posterior ensemble retains sufficient
distributional flexibility to improve its joint alignment with the empirical
population-moment distributions. At larger $M$, departures from the
prior-implied ensemble distribution become increasingly costly. The
RR model's inability to reconcile $[\Sigma_e]_{11}$ with the other
four moments then becomes dominant, and the generalized evidence declines.
The location of the turning point and the subsequent decline therefore
diagnose the degree of tension between the structural model and the declared
empirical target, conditional on the structural prior and the other
target-defining quantities.

When the structural restrictions are compatible with the declared empirical
target, this reversal need not occur. In the correctly specified FR
benchmark reported in Online Appendix~D, the average generalized evidence
increases monotonically over the same grid of $M$. The contrast between the
two profiles supports the interpretation of the decline in the RR
model as a consequence of the omitted-shock restriction rather than a generic
feature of the DMPI computation.

Online Appendix~D also reports a classical minimum-distance (CMD) benchmark using
the same datasets and treating their sample moments as point targets. Its
criterion is small for the correctly
specified FR model but large for the RR model in all 30
replications. Under its conventional specification-test interpretation, CMD
asks whether the exact structural restrictions are accepted or rejected. The hump-shaped evidence profile of DMPI shows that the
RR restrictions remain useful when imposed softly, even though they fail as
an exact representation of the sample moments. The appendix also explains why
SMM or indirect inference based
on the same unrestricted VAR auxiliary estimator is not well defined under
the stochastically singular RR model.

Accordingly, the evidence-maximizing value of $M$ identifies, among the
prespecified values considered, the degree of structural discipline most
strongly supported by the declared population-moment distributions. This
role is analogous to the selection of the DSGE-prior tightness parameter
$\lambda$ by marginal likelihood in \citet{delnegroschrfheide04}.

\vspace{1mm}
\begin{center}
{\large \textbf{4. DMPI in a three-equation New Keynesian DSGE model}}
\end{center}

\vspace{1mm}
This section applies DMPI to postwar U.S. data in a three-equation New
Keynesian setting. Using an unrestricted VAR as the empirical model $E$, we
compare an FR structural model with stochastically singular RR alternatives
against a common empirical target.

\vspace{1mm}
\noindent
\textit{4.1. The FR and RR NK--DSGE models}

\vspace{1mm}
We use the three-equation NK--DSGE model studied by
\citet{delnegroschrfheide04}. Let $\tilde{x}_t$, $\tilde{\pi}_t$, and $\tilde{R}_t$
denote the log deviations of detrended output, gross inflation, and the gross
nominal interest rate from their steady-state values. The log-linearized
equilibrium conditions are
\begin{align*}
\tilde{x}_t
&=\mathbf E_t\tilde{x}_{t+1}
-\tau^{-1}\left(\tilde{R}_t-\mathbf E_t\tilde{\pi}_{t+1}\right)
+(1-\rho_g)\tilde{g}_t+\rho_z\tau^{-1}\tilde{z}_t, \\
\tilde{\pi}_t
&=\frac{\gamma}{r^*}\mathbf E_t\tilde{\pi}_{t+1}
+\kappa(\tilde{x}_t-\tilde{g}_t), \\
\tilde{R}_t
&=\rho_R\tilde{R}_{t-1}
+(1-\rho_R)(\psi_1\tilde{\pi}_t+\psi_2\tilde{x}_t)
+\varepsilon_{R,t}.
\end{align*}
Here $\mathbf E_t$ denotes expectation conditional on information available
at time $t$; $\tau$ is the inverse intertemporal elasticity of substitution;
$\gamma$ and $r^*$ are the gross deterministic growth rate and steady-state
gross real interest rate; $\kappa$ is the slope of the NKPC; and
$\rho_R$, $\psi_1$, and $\psi_2$ govern interest-rate smoothing and the policy
responses to inflation and detrended output.

The FR model contains technology-growth, government-spending, and
monetary-policy innovations. The first two states satisfy
$\tilde z_t=\rho_z\tilde z_{t-1}+\varepsilon_{z,t}$ and
$\tilde g_t=\rho_g\tilde g_{t-1}+\varepsilon_{g,t}$, where $\rho_z$ and
$\rho_g$ are persistence parameters. The innovations are mutually independent,
with $\varepsilon_{j,t}\stackrel{\mathrm{i.i.d.}}{\sim}
\mathcal N(0,\sigma_j^2)$ for $j\in\{z,g,R\}$.

The measurement equations are
\begin{align*}
\Delta\ln x_t
&=\ln\gamma+\Delta\tilde{x}_t+\tilde{z}_t, \\
\Delta\ln P_t
&=\ln\pi^*+\tilde{\pi}_t, \\
\ln R_t^a
&=4\left(\ln r^*+\ln\pi^*+\tilde{R}_t\right),
\end{align*}
where $\pi^*$ is steady-state gross inflation and $R_t^a$ is the annualized
gross nominal interest rate. The observable vector is
$\mathbf y_t=(\Delta\ln x_t,\Delta\ln P_t,\ln R_t^a)'$.

We consider three RR specifications, indexed by the innovation that
is suppressed. The specification RR\textsubscript{g} imposes
$\tilde g_t\equiv0$ and removes $\rho_g$ and $\sigma_g$ from the structural
parameter vector. The specification RR\textsubscript{z} imposes
$\tilde z_t\equiv0$ and removes $\rho_z$ and $\sigma_z$. Finally,
RR\textsubscript{R} sets
$\varepsilon_{R,t}\equiv0$ and removes $\sigma_R$; the monetary-policy rule,
including the smoothing parameter $\rho_R$, remains part of the model. Thus,
each reduced-rank specification retains two structural shocks for three
observables. Its spectral density matrix for $\mathbf y_t$ has rank at most
two, so the usual nonsingular Gaussian full-information likelihood is
unavailable without measurement error or an auxiliary shock. For every
structurally admissible parameter value, however, the population-moment
mappings used below remain finite, and the DMPI target is well defined. The
three RR specifications allow us to examine whether model comparison depends on
which structural innovation is omitted, rather than on rank alone.

\vspace{1mm}
\noindent
\textit{4.2. The unrestricted VAR as the empirical model}

\vspace{1mm}
As in \citet{delnegroschrfheide04}, the empirical model $E$ is an
unrestricted VAR(4) for the observable vector $\mathbf y_t$:
\begin{equation*}
\mathbf y_t
=c+\sum_{\ell=1}^{4}A_\ell\mathbf y_{t-\ell}+\mathbf e_t,
\qquad
\mathbf e_t\stackrel{\mathrm{i.i.d.}}{\sim}
\mathcal N(\mathbf 0,\Sigma_e),
\end{equation*}
where $c$ is a vector of constants, $A_\ell$ are coefficient matrices, and
$\Sigma_e$ is a positive-definite covariance matrix.

The conventional DSGE-VAR uses DSGE-implied population moment matrices to
construct a prior over the VAR coefficients and innovation covariance matrix.
For a chosen lag order, the DSGE-implied population moment matrices determine
the corresponding population linear projection. They can be represented as
cross-products of dummy observations and used to center the VAR prior; the
construction does not require the DSGE observables themselves to follow an
exact finite-order VAR. The resulting prior thereby imposes structural
discipline softly across the full VAR parameterization.

DMPI instead uses the unrestricted VAR only to quantify uncertainty about a
declared set of population moments. For each selected moment, the DSGE-implied
ensemble values enter the DM construction as structural cell counts, thereby
imposing structural discipline softly at the level of its distribution. Thus,
both approaches use DSGE-implied quantities as soft structural discipline, but
DSGE-VAR applies it across the full VAR parameterization, whereas DMPI applies
it directly to the selected population-moment distributions.

Before comparing the four specifications, we fix 21 population moments: three means, three variances, three
contemporaneous covariances, and four own autocovariances for each observable. These moments summarize
long-run averages, unconditional volatility, contemporaneous comovement, and
persistence--the principal dimensions of business-cycle dynamics that the
NK-DSGE model is intended to explain. We do not target higher-lag
population cross-covariances, such as
$\operatorname{Cov}(\Delta\ln x_t,\ln R^a_{t-4})$, which may be informative
for other purposes but fall outside the declared scope of this application.
The positive definiteness of $\Sigma_e$ is a property of the empirical model
$E$ and imposes no corresponding rank restriction on the structural model
$A$. The FR model and all three RR specifications can therefore be mapped
directly into the same population moments.

\vspace{1mm}
\noindent
\textit{4.3. Data, target construction, and computation}

\vspace{1mm}
The quarterly sample spans 1954:Q4--2001:Q3 and contains output growth,
inflation, and the annualized federal funds rate.\footnote{The series are
GDPC96, CPIAUCSL, and FEDFUNDS from the Federal Reserve Bank of St. Louis
FRED database. Output growth and inflation are measured as 100 times the
quarterly log difference; the federal funds rate is expressed in annualized
percentage points.} We obtain $N_{\mathrm{MC}}=50{,}000$ independent draws
by direct simulation from the conjugate normal--inverse-Wishart posterior
$p(\theta_E\mid\mathbf y,E)$ and transform every draw into the 21 population
moments specified in Section~4.2. The posterior uncertainty of the empirical
model, including the sampling uncertainty associated with the finite data
sample, is therefore propagated into the empirical population-moment
distributions.

For each moment, the interval between the empirical 0.5th and 99.5th
percentiles is divided into 100 equal-width cells and complemented by lower
and upper tail cells, giving $K=102$ cells in total. The
$N_{\mathrm{MC}}$ draws determine both the partition endpoints and the
empirical cell probabilities. We set
$N_{\mathrm{eff}}=N_{\mathrm{MC}}=50{,}000$ and apply the
largest-remainder rule, which here leaves the integer cell counts unchanged.
This target construction avoids an additional coarsening of the simulated
empirical population-moment distributions. We set $\delta=1$. The resulting
count vectors and partitions are fixed before structural inference and held
common across the FR model, the three RR specifications, and all values of
$M$.

Table~\ref{tab:prior_dsge_var} summarizes the independent base priors for the
FR model. Their centers follow the calibration in
\citet{delnegroschrfheide04}, with the dispersions reported in the table. For
each RR specification, we remove the inactive parameters from the FR base
prior. For every structural specification $A$, we condition the resulting
base prior on that specification's structural admissibility set to obtain $\pi_A$ in
equation~(\ref{eq:conditional_prior}); its $M$-fold product is the ensemble
prior in equation~(\ref{direct_prior}). The FR model has 13 parameters,
RR\textsubscript{g} and RR\textsubscript{z} have 11 each, and
RR\textsubscript{R} has 12.

\begin{table}[!htbp]
\centering
\begin{minipage}{0.64\textwidth}
\captionsetup{font=footnotesize}
\caption{Base priors for the FR structural parameters}
\label{tab:prior_dsge_var}

\centering
\scriptsize
\renewcommand{\arraystretch}{0.7}
\begin{tabular}{lccc}
\hline
Parameter & Distribution & Center & Dispersion \\
\hline
$\ln \gamma$  & Normal           & 0.500 & 0.125 \\
$\ln \pi^*$   & Normal           & 1.000 & 0.250 \\
$\ln r^*$     & Gamma            & 0.500 & 0.125 \\
$\kappa$      & Beta             & 0.300 & 0.075 \\
$\tau$        & Gamma            & 2.000 & 0.250 \\
$\psi_1$      & Gamma            & 1.500 & 0.125 \\
$\psi_2$      & Gamma            & 0.125 & 0.050 \\
$\rho_R$      & Beta             & 0.500 & 0.100 \\
$\rho_g$      & Beta             & 0.800 & 0.050 \\
$\rho_z$      & Beta             & 0.300 & 0.050 \\
$\sigma_R$    & Folded Normal    & 0.251 & 0.075 \\
$\sigma_g$    & Folded Normal    & 0.630 & 0.075 \\
$\sigma_z$    & Folded Normal    & 0.875 & 0.050 \\
\hline
\end{tabular}

\par\vspace{1mm}
\raggedright
\scriptsize
\textit{Note.} Parameters $\ln\gamma$, $\ln\pi^*$, and $\ln r^*$ are
expressed in percentage terms. Except in the Folded Normal rows, center and
dispersion denote the prior mean and standard deviation. For each shock
standard deviation, $\sigma_j=|s_j|$ with
$s_j\sim\mathcal N(\mu_j,\tau_j^2)$; the table reports the underlying
location $\mu_j$ and scale $\tau_j$. The RR specifications inherit the
relevant rows: RR\textsubscript{g} omits $\rho_g$ and $\sigma_g$,
RR\textsubscript{z} omits $\rho_z$ and $\sigma_z$, and RR\textsubscript{R} omits
$\sigma_R$.
\end{minipage}
\end{table}
We evaluate the FR and RR\textsubscript{g} models at
$M\in\{1,2,5,10,20,30,50\}$ and compare all four structural specifications at
$M\in\{1,2,10,50\}$. All calculations use the exact
DM score and the annealed SMC procedure in Section~2.6.\footnote{Preliminary
calculations determine fixed model-specific annealing schedules. We apply 12 mutation
sweeps at each annealing stage and report estimates at the larger of two
particle counts, pooling two independent runs. The reported FR estimates use
$P=8192$ for $M\in\{1,2,5\}$ and $P=4096$ for
$M\in\{10,20,30,50\}$. The three RR specifications use $P=1024$. Numerical
diagnostics include seed dispersion, particle-count comparisons,
incremental-weight ESS, and maximum particle weights.}

Online Appendix~E collects target sensitivities, numerical diagnostics,
and the RR posterior tables and distributional figures.

\vspace{1mm}
\noindent
\textit{4.4. Empirical results}

\vspace{1mm}
\noindent
\textit{4.4.1. Posterior inference and distributional fit in the FR model}

\vspace{1mm}
\suppressfloats[t]

\begin{figure}[H]
\centering
\includegraphics[width=0.91\textwidth]
{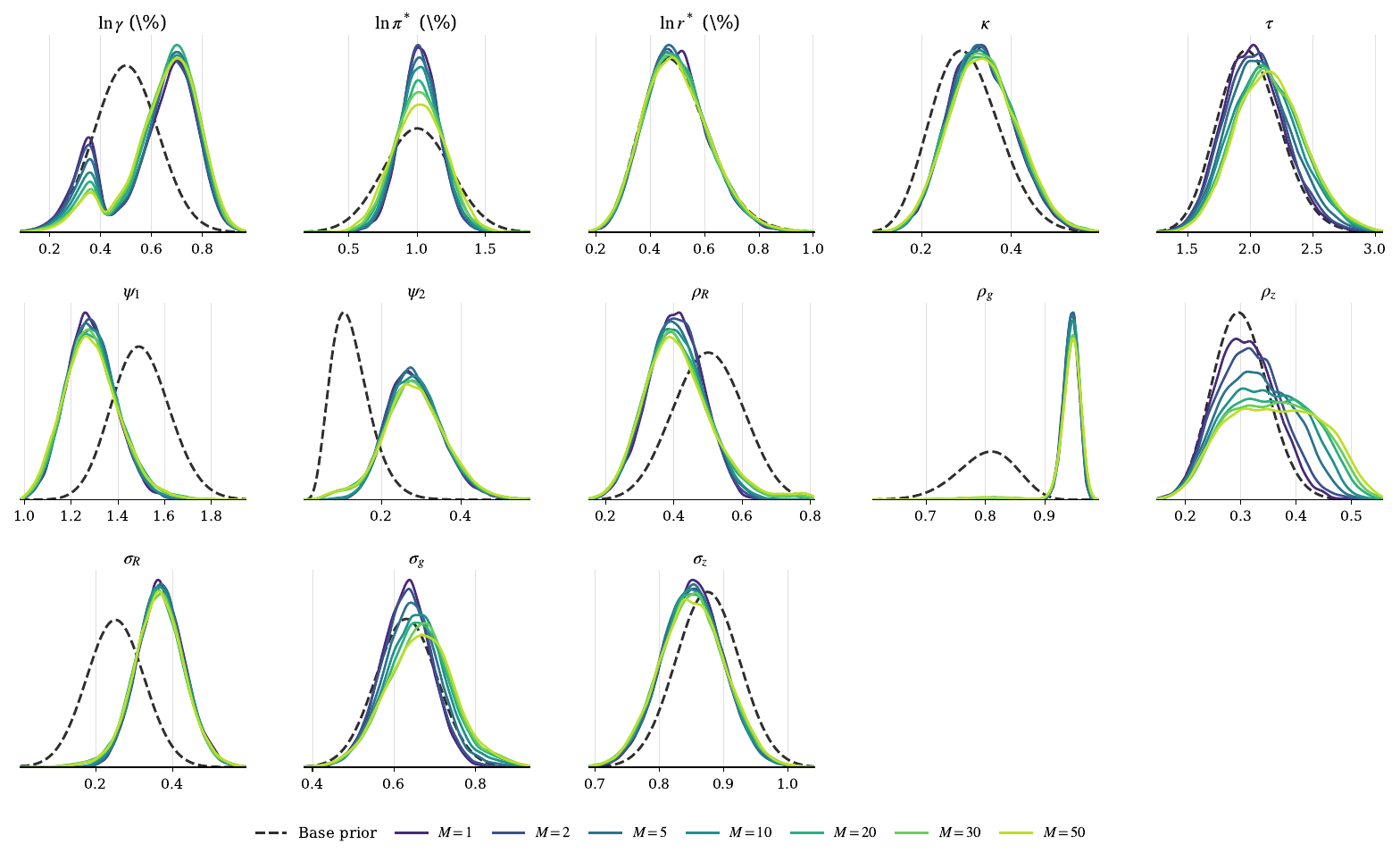}
\captionsetup{font=footnotesize}
\caption{Posterior distributions of the structural parameters in the FR
model. Solid curves show the marginal posterior densities for
$M\in\{1,2,5,10,20,30,50\}$, and dashed curves show the corresponding base-prior
densities. The posterior densities use the exchangeable ensemble draws from
the final SMC particles across two independent runs. The parameters
$\ln\gamma$, $\ln\pi^*$, and $\ln r^*$ are expressed in percentage terms;
the shock standard deviations are plotted as $\sigma_j=|s_j|$.}
\label{fig:us_fr_parameter_kde}
\end{figure}

For the FR model, Table~\ref{tab:posterior_fr_data} reports posterior summaries
and the log joint generalized evidence, while Figure~\ref{fig:us_fr_parameter_kde}
displays the corresponding parameter distributions.

\begin{table}[!htbp]
\centering
\captionsetup{font=footnotesize}
\caption{Posterior inference and generalized evidence for the FR model}
\label{tab:posterior_fr_data}
\scriptsize
\setlength{\tabcolsep}{2.0pt}
\renewcommand{\arraystretch}{0.72}
\begin{adjustbox}{max width=\textwidth}
\begin{tabular}{llccccccc}
\hline
Parameter & Prior mean & $M=1$ & $M=2$ & $M=5$ & $M=10$
& $M=20$ & $M=30$ & $M=50$ \\
\hline
$\ln\gamma$ (\%) & 0.500
& 0.599 & 0.604 & 0.617 & 0.628 & 0.637 & 0.640 & 0.642 \\
& & [0.250, 0.837] & [0.253, 0.840] & [0.259, 0.843]
& [0.272, 0.847] & [0.284, 0.850] & [0.289, 0.853] & [0.297, 0.859] \\
$\ln\pi^*$ (\%) & 1.000
& 1.018 & 1.018 & 1.019 & 1.017 & 1.017 & 1.017 & 1.017 \\
& & [0.732, 1.302] & [0.731, 1.305] & [0.724, 1.313]
& [0.710, 1.328] & [0.687, 1.348] & [0.669, 1.364] & [0.644, 1.388] \\
$\ln r^*$ (\%) & 0.500
& 0.498 & 0.499 & 0.500 & 0.500 & 0.499 & 0.499 & 0.500 \\
& & [0.295, 0.743] & [0.297, 0.746] & [0.295, 0.750]
& [0.293, 0.755] & [0.292, 0.758] & [0.290, 0.759] & [0.289, 0.763] \\
$\kappa$ & 0.300
& 0.332 & 0.334 & 0.337 & 0.338 & 0.338 & 0.338 & 0.339 \\
& & [0.201, 0.479] & [0.202, 0.483] & [0.202, 0.485]
& [0.202, 0.489] & [0.200, 0.491] & [0.199, 0.493] & [0.198, 0.494] \\
$\tau$ & 2.000
& 2.032 & 2.047 & 2.086 & 2.120 & 2.145 & 2.156 & 2.165 \\
& & [1.591, 2.534] & [1.594, 2.570] & [1.615, 2.637]
& [1.632, 2.680] & [1.642, 2.703] & [1.651, 2.722] & [1.655, 2.732] \\
$\psi_1$ & 1.500
& 1.285 & 1.287 & 1.287 & 1.288 & 1.293 & 1.289 & 1.287 \\
& & [1.090, 1.502] & [1.093, 1.506] & [1.091, 1.507]
& [1.088, 1.515] & [1.086, 1.550] & [1.082, 1.545] & [1.077, 1.548] \\
$\psi_2$ & 0.125
& 0.284 & 0.284 & 0.287 & 0.286 & 0.282 & 0.283 & 0.283 \\
& & [0.164, 0.431] & [0.165, 0.432] & [0.165, 0.438]
& [0.159, 0.439] & [0.119, 0.442] & [0.121, 0.443] & [0.116, 0.446] \\
$\rho_R$ & 0.500
& 0.405 & 0.404 & 0.404 & 0.406 & 0.409 & 0.411 & 0.414 \\
& & [0.249, 0.554] & [0.249, 0.558] & [0.247, 0.572]
& [0.246, 0.583] & [0.245, 0.612] & [0.244, 0.627] & [0.243, 0.663] \\
$\rho_g$ & 0.800
& 0.944 & 0.945 & 0.944 & 0.944 & 0.939 & 0.939 & 0.939 \\
& & [0.917, 0.967] & [0.918, 0.967] & [0.917, 0.968]
& [0.915, 0.968] & [0.808, 0.969] & [0.813, 0.969] & [0.807, 0.969] \\
$\rho_z$ & 0.300
& 0.308 & 0.316 & 0.330 & 0.343 & 0.353 & 0.358 & 0.363 \\
& & [0.208, 0.416] & [0.209, 0.432] & [0.211, 0.457]
& [0.213, 0.474] & [0.215, 0.492] & [0.216, 0.501] & [0.216, 0.512] \\
$\sigma_R$ & 0.251
& 0.371 & 0.369 & 0.369 & 0.368 & 0.363 & 0.364 & 0.364 \\
& & [0.255, 0.494] & [0.253, 0.492] & [0.253, 0.493]
& [0.248, 0.492] & [0.227, 0.492] & [0.229, 0.492] & [0.227, 0.494] \\
$\sigma_g$ & 0.630
& 0.629 & 0.632 & 0.642 & 0.652 & 0.659 & 0.663 & 0.665 \\
& & [0.504, 0.749] & [0.502, 0.758] & [0.503, 0.782]
& [0.505, 0.811] & [0.504, 0.828] & [0.503, 0.834] & [0.500, 0.834] \\
$\sigma_z$ & 0.875
& 0.852 & 0.852 & 0.851 & 0.850 & 0.851 & 0.851 & 0.852 \\
& & [0.760, 0.946] & [0.759, 0.946] & [0.757, 0.945]
& [0.754, 0.947] & [0.753, 0.950] & [0.752, 0.950] & [0.752, 0.952] \\
\hline
Posterior average $\log L_M$ & --
& -15187.21 & -15161.18 & -15093.12 & -15002.80
& -14884.29 & -14803.92 & -14732.48 \\
Implied KL cost & --
& 17.69 & 35.41 & 90.33 & 182.63
& 358.93 & 542.87 & 906.75 \\
$\log \widehat Z_{\mathrm{FR},M}$ & --
& -15204.89 & -15196.59 & -15183.45 & -15185.43
& -15243.21 & -15346.79 & -15639.22 \\
\hline
\end{tabular}
\end{adjustbox}

\vspace{1mm}
\begin{minipage}{0.96\textwidth}
\scriptsize
\textit{Note.} Posterior means are reported in the first row and 95\%
credible intervals in the second row. The posterior summaries pool the
exchangeable structural-parameter draws from the final SMC particles across
the two independent runs used at the reported particle count. The parameters
$\ln\gamma$, $\ln\pi^*$, and $\ln r^*$ are expressed in percentage terms.
For the shock standard deviations, the summaries refer to
$\sigma_j=|s_j|$ as defined in Table~\ref{tab:prior_dsge_var}. The final row
reports the estimated log joint generalized evidence. The posterior average
log DM score is $\mathbb E_{\Pi_{\mathrm{FR},M}}[\log L_M]$; the implied
posterior-to-prior KL cost is recovered residually from the evidence identity
given below.
\end{minipage}
\end{table}

The selected population moments update some parameters materially while
leaving others close to their prior locations. Across the examined values of
$M$, the posterior means of $\psi_1$, $\psi_2$, $\rho_g$, and $\sigma_R$
remain displaced from their prior means in essentially the same directions:
$\psi_1$ is lower, $\psi_2$ and $\sigma_R$ are higher, and $\rho_g$ is
concentrated at high persistence. These four means vary only modestly with
$M$. Other parameters move more gradually, most notably $\tau$, $\rho_z$, and
$\sigma_g$. The credible intervals do not contract mechanically with $M$ and
in several cases widen.

Figure~\ref{fig:us_fr_parameter_kde} shows distributional changes that are not
fully summarized by posterior means and credible intervals. The posterior of
$\ln\gamma$ is robustly bimodal: as $M$ rises, the relative mass below 0.5
falls from 0.262 to 0.149, although both modes remain visible. The posterior
of $\rho_g$ stays sharply concentrated near 0.95 but develops a longer
low-persistence tail, while the distribution of $\rho_z$ shifts to the right
and develops a shoulder. By contrast, the posteriors of $\psi_2$ and $\rho_R$
are unimodal. These results illustrate that
$M$ can alter tails and relative mass across parameter regions even when the
posterior mean changes little.

\begin{figure}[H]
\centering
\includegraphics[width=0.97\textwidth]
{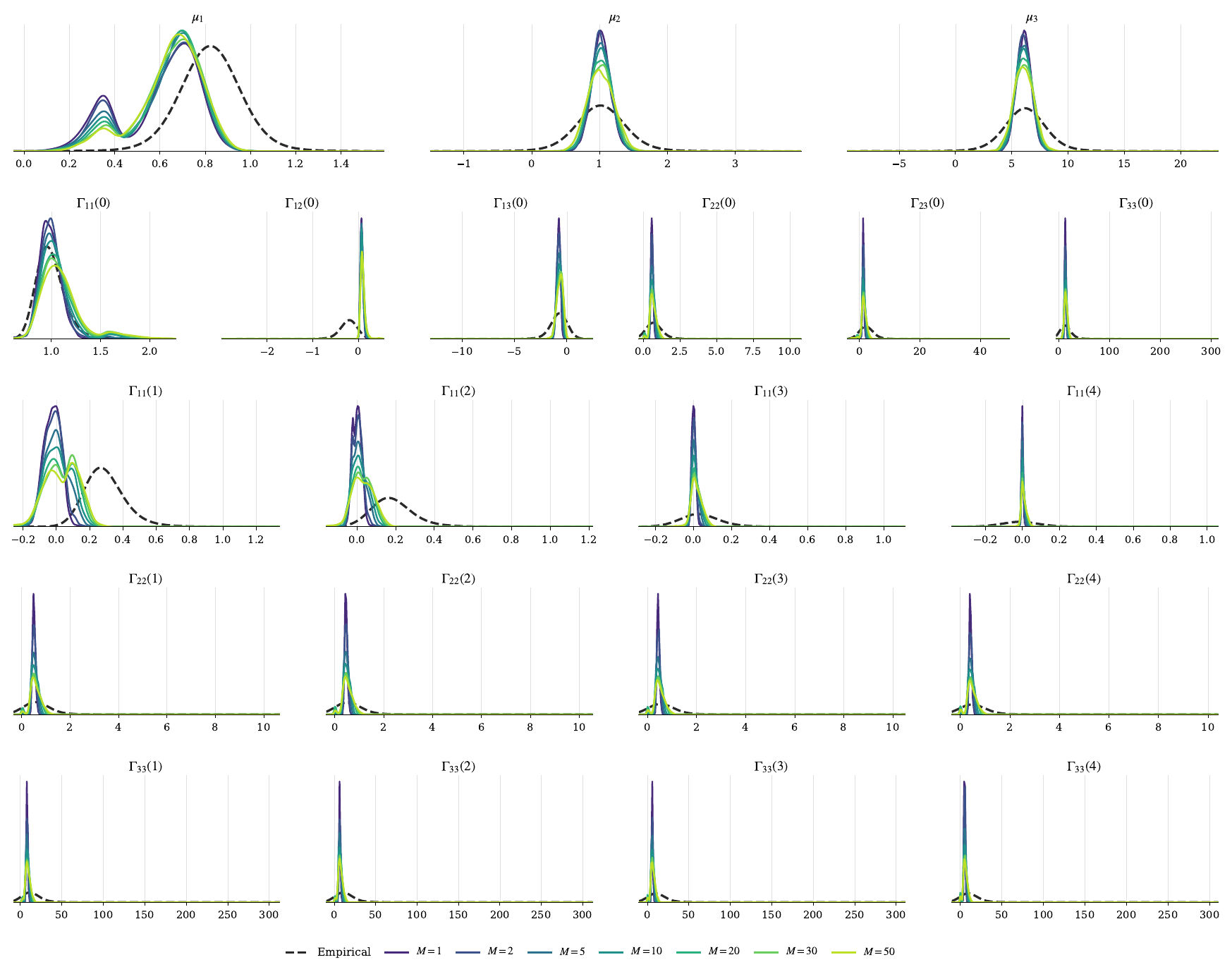}
\captionsetup{font=footnotesize}
\caption{Empirical and posterior model-implied distributions of the 21
selected population moments under the FR model. Here
$\mu_i\equiv\mathbb E(y_{i,t})$ and
$\Gamma_{ij}(h)\equiv\operatorname{Cov}(y_{i,t},y_{j,t-h})$. The dashed curve
in each panel is the empirical posterior distribution, and the solid curves
are the model-implied distributions for
$M\in\{1,2,5,10,20,30,50\}$. The latter are computed from the exchangeable
structural-parameter draws in the final SMC particles.}
\label{fig:us_fr_moment_distributions}
\end{figure}

Figure~\ref{fig:us_fr_moment_distributions} shows that the quality of the
distributional match varies substantially across the 21 population moments. The FR
model produces considerable overlap with the empirical
distributions of $\mu_2$, $\mu_3$, and $\Gamma_{11}(0)$, as well as partial
overlap for several other contemporaneous and dynamic moments. The most
systematic discrepancy concerns output growth. Although the model broadly
reproduces its unconditional variance, it understates mean output growth and
generates too little short-run persistence. The bimodality of the
model-implied distribution of $\mu_1$ is inherited from the two modes of
$\ln\gamma$ shown in Figure~\ref{fig:us_fr_parameter_kde}. In particular, the empirical
distributions of $\Gamma_{11}(1)$ and $\Gamma_{11}(2)$ are centered at
positive values, whereas their model-implied counterparts remain concentrated
near zero. A clear discrepancy also appears for the contemporaneous covariance
$\Gamma_{12}(0)$. In addition, the model-implied distributions of several
inflation and interest-rate variance and autocovariance moments are more
concentrated than their empirical counterparts. Hence, a full-rank shock
structure removes stochastic singularity but does not by itself ensure that
the plain-vanilla NK model reproduces the means, comovement, and persistence
of the selected U.S. population moments.

To distinguish posterior average fit from the cost of departing from the
product prior, we use the exact identity
\[
\log Z_{A,M}
=
\mathbb E_{\Pi_{A,M}}[\log L_M]
-
D_{\mathrm{KL}}\!\left(
\Pi_{A,M}\Vert\Pi_{A,M}^{0}
\right).
\]
Here $D_{\mathrm{KL}}(\Pi_{A,M}\Vert\Pi_{A,M}^{0})$ is the
Kullback--Leibler divergence of the generalized posterior from the
product prior.
The first term measures posterior average fit, whereas the second measures
the extent to which the generalized posterior must be reweighted away from
the model-specific product prior. This is a posterior-to-prior
KL cost, not a decomposition into separate likelihood and prior-density
scores. We recover the implied KL cost residually from this identity. The
independent numerical check used below is that the reconstructed 21 log DM
scores sum to the log DMPI score stored by the SMC calculation for every
posterior particle.

The final row of Table~\ref{tab:posterior_fr_data} reports the joint
generalized evidence associated with these patterns of distributional fit.
The estimated log evidence rises from
$-15204.89$ at $M=1$ to $-15183.45$ at $M=5$, declines by 1.98 log points at
$M=10$, and then falls sharply to $-15639.22$ at $M=50$. Thus, among the
prespecified values, the evidence-maximizing discipline is
$\widehat M_{\mathrm{FR}}=5$. Applying the identity above, the posterior
average log DMPI score improves monotonically
from $-15187.21$ at $M=1$ to $-14732.48$ at $M=50$, but its marginal
improvement diminishes. The implied posterior-to-prior KL cost rises from
17.69 to 906.75 and is nearly constant per ensemble member. The maximum at
$M=5$ is therefore consistent with the point at which the diminishing
improvement in the posterior average log DMPI score is overtaken by the
rising reweighting cost.

\vspace{1mm}
\noindent
\textit{4.4.2. Evidence across RR specifications}

\vspace{1mm}
Table~\ref{tab:evidence_shock_specs} compares the FR model with all three RR
specifications at the common values $M\in\{1,2,10,50\}$. Because the empirical
counts, partitions, $N_{\mathrm{eff}}$, and smoothing parameter are identical
across models, the joint generalized evidences are directly comparable at
each value of $M$.

\begin{table}[H]
\centering
\begin{minipage}{0.88\textwidth}
\captionsetup{font=footnotesize}
\caption{Generalized-evidence differences relative to the FR model}
\label{tab:evidence_shock_specs}

\centering
\small
\setlength{\tabcolsep}{7pt}
\renewcommand{\arraystretch}{0.88}
\begin{tabular}{lccccc}
\hline
Model & Innovations retained & $M=1$ & $M=2$ & $M=10$ & $M=50$ \\
\hline
FR              & $z,g,R$ & 0.00 & 0.00 & 0.00 & 0.00 \\
RR\textsubscript{g}          & $z,R$   & $-12.18$ & $-26.46$ & $-158.48$ & $-890.01$ \\
RR\textsubscript{z}          & $g,R$   & $\mathbf{4.53}$ & $\mathbf{8.30}$ & $27.94$ & $31.01$ \\
RR\textsubscript{R}          & $g,z$   & $3.02$ & $7.14$ & $\mathbf{54.43}$ & $\mathbf{327.35}$ \\
\hline
\end{tabular}

\par\vspace{1mm}
\raggedright
\scriptsize
\textit{Note.} Entries are
$\widehat\Delta_M(A,\mathrm{FR})
=\log\widehat Z_{A,M}-\log\widehat Z_{\mathrm{FR},M}$, so a positive value
favors the model in that row. Boldface identifies the largest generalized
evidence at each $M$. The estimates use the validated two-run comparison
cohorts; model-specific annealing schedules are used for RR\textsubscript{z}
and RR\textsubscript{R}.
\end{minipage}
\end{table}

The ranking is not determined by stochastic rank alone. The
RR\textsubscript{g} model, which omits the government-spending innovation, has
lower generalized evidence than the FR model at all four common values of $M$.
By contrast, both RR\textsubscript{z} and RR\textsubscript{R} exceed the FR model
throughout the comparison. Among the four specifications,
RR\textsubscript{z} has the largest evidence at $M=1$ and $M=2$, while
RR\textsubscript{R} has the largest evidence at $M=10$ and $M=50$.

The comparisons suggest that retaining the government-spending innovation is
particularly important for compatibility with the selected population
moments under the priors used here. All three RR specifications remove
the prior-predictive variation associated with one innovation;
RR\textsubscript{g} and RR\textsubscript{z} also remove the corresponding
persistence parameter, whereas RR\textsubscript{R} retains the monetary-policy
rule and $\rho_R$. The evidence differences therefore compare the structural
specifications as a whole rather than innovation variances in isolation. Over the four
values considered, the absolute evidence gaps widen as $M$ increases. Stronger structural discipline therefore makes the differences
among the shock specifications more visible.

To identify which population moments account for model differences in the
posterior average log DMPI score, define the posterior average log DM score
$\bar s_{i,M}^{A}=\mathbb E_{\Pi_{A,M}}[\log L_{i,M}]$. Table~\ref{tab:rrR_fr_block_fit} reports
$d_{i,M} =
\bar s_{i,M}^{\mathrm{RR}_{R}}-\bar s_{i,M}^{\mathrm{FR}}$. Positive entries indicate a better posterior average fit under
RR\textsubscript{R} for that moment. Because
$\log L_M=\sum_{i=1}^{I}\log L_{i,M}$,
the sum $\sum_{i=1}^{I}d_{i,M}$ equals the difference in posterior average
fit between RR\textsubscript{R} and FR.

At $M=2$, gains and losses across moments nearly offset, and the posterior
average score is 3.96 log points lower under RR\textsubscript{R}. Its implied
posterior-to-prior KL cost is nevertheless 11.09 log points smaller than that
of FR. The lower reweighting cost more than offsets the modest fit disadvantage
and produces the 7.14-log-point evidence advantage in
Table~\ref{tab:evidence_shock_specs}. The same mechanism operates at $M=10$:
the average-score difference is $-5.47$, whereas the KL-cost difference is
$-59.91$, yielding an evidence advantage of 54.43 log points. At $M=50$,
RR\textsubscript{R} also has a 28.61-log-point advantage in posterior average
fit, while its KL cost is 298.74 log points smaller.

\begin{table}[H]
\centering
\captionsetup{font=footnotesize}
\caption{Principal population-moment fit differences: RR\textsubscript{R}
relative to FR}
\label{tab:rrR_fr_block_fit}
\scriptsize
\setlength{\tabcolsep}{7pt}
\renewcommand{\arraystretch}{0.82}
\begin{tabular}{lrrrr}
\hline
Moment & $M=1$ & $M=2$ & $M=10$ & $M=50$ \\
\hline
$\Gamma_{12}(0)$ & $0.68$ & $2.05$ & $19.84$ & $137.94$ \\
$\Gamma_{11}(1)$ & $1.16$ & $2.87$ & $19.48$ & $105.13$ \\
$\Gamma_{11}(2)$ & $1.14$ & $2.11$ & $8.39$ & $35.50$ \\
$\Gamma_{13}(0)$ & $-2.96$ & $-5.90$ & $-30.09$ & $-151.10$ \\
$\Gamma_{11}(0)$ & $-1.30$ & $-2.63$ & $-13.26$ & $-64.53$ \\
\textit{All 21 moments} & $-2.56$ & $-3.96$ & $-5.47$ & $28.61$ \\
\hline
\end{tabular}

\vspace{1mm}
\begin{minipage}{0.94\textwidth}
\scriptsize
\textit{Note.} Entries are differences in posterior average log
DM scores, $d_{i,M}=\bar
s_{i,M}^{\mathrm{RR}_{R}}-\bar s_{i,M}^{\mathrm{FR}}$, averaged equally over
two independent runs. Positive values favor RR\textsubscript{R}. The five
moments shown have the largest absolute contributions at $M=50$; the final
row sums all 21 moments. Because all moments are evaluated using each model's
common structural ensemble, the entries are fit diagnostics rather than
moment-specific generalized evidences. The complete decomposition is reported in
Online Appendix~E.
\end{minipage}
\end{table}

The moment-specific posterior average log DM scores show that
RR\textsubscript{R} fits some moments better and others worse than FR. The
principal gains for RR\textsubscript{R} occur in the
contemporaneous output-growth--inflation covariance, $\Gamma_{12}(0)$, and
the first two output-growth autocovariances, $\Gamma_{11}(1)$ and
$\Gamma_{11}(2)$, precisely where Figure~
\ref{fig:us_fr_moment_distributions} reveals important FR discrepancies.
These gains are partly offset by poorer fit for the variance of output growth,
$\Gamma_{11}(0)$, and its contemporaneous covariance with the interest rate,
$\Gamma_{13}(0)$. The other RR
specifications reinforce the distinction between average fit and evidence.
RR\textsubscript{g} has a smaller KL cost than FR, but its much poorer average
fit more than offsets that advantage. For RR\textsubscript{z}, and for
RR\textsubscript{R} at $M\leq10$, the evidence advantage instead arises despite
a lower posterior average log DMPI score because the implied KL cost is
sufficiently smaller. Thus, the reduced-rank rankings reflect both model-specific fit and
the amount of posterior reweighting required under each structural prior. The
principal RR\textsubscript{R} ranking is preserved in prespecified robustness
exercises.\footnote{RR\textsubscript{R} remains ahead of FR at every common
$M$ in all robustness checks. See Online Appendix Tables~E.5--E.7.}

\vspace{1mm}
\begin{center}
{\large \textbf{5. Concluding remarks}}
\end{center}

This paper has developed DMPI, a generalized Bayesian framework that extends
the MEI of \citet{geweke10} to posterior inference and model comparison for
incomplete structural models. DMPI compares structural and empirical
distributions of selected population moments using an exact DM score, while
the joint generalized-evidence profile over
$M$ shows how empirical support changes with structural discipline.

More broadly, DMPI permits posterior inference and model comparison without
augmenting a structural model with auxiliary stochastic assumptions added
solely for econometric convenience. It thereby preserves the model's intended
explanatory scope and allows its theoretical restrictions to be assessed and
imposed at an empirically supported strength.

In the NKPC experiments, a stochastically singular RR model that
omits a shock present in the data-generating process has a hump-shaped
generalized-evidence profile: moderate structural discipline raises the
evidence, whereas stronger discipline exposes the omitted-shock restriction.
Thus, a model may fail as an exact representation while retaining useful
theoretical restrictions when they are imposed softly.

The U.S. application of the three-equation NK-DSGE model shows that the
effect of restoring full rank depends on which structural innovation is
added: adding the government-spending
innovation raises generalized evidence, whereas adding either the technology
or monetary-policy innovation lowers it at common values of $M$. Most notably, RR\textsubscript{R}, which omits the monetary-policy
innovation, has higher evidence than the full-rank model at every common $M$
and the highest evidence among the four specifications at the moderate and
larger values considered. This finding questions whether a Taylor-rule
disturbance is an appropriate generator of unconditional business-cycle
fluctuations, while leaving intact its role in studying the propagation of
monetary-policy shocks.

One promising extension is to use local projections as a limited-information
empirical reference model for posterior distributions of population impulse
responses. The generalized-evidence profile over $M$ would then assess the
empirically supported strength of the structural model's restrictions on
those responses. Other applications include equilibrium asset-pricing models
and heterogeneous-agent or other nonlinear DSGE models in which distributions
are central theoretical objects.

\renewcommand{\baselinestretch}{1}
\singlespacing

%\clearpage
\bibliography{dmpi_0926_rr}

\input{appendix_dmpi_0926_rr.tex}

\end{document}

%% file: appendix_dmpi_0926_rr.tex
% Included after the bibliography by dmpi_0926_rr.tex for arXiv.
\clearpage
\appendix
\section*{Online Appendix}
\pagestyle{plain}
\onehalfspacing

\noindent
\textbf{Appendix A: Exact Dirichlet--multinomial construction}

\vspace{3mm}
This appendix derives equation~(8) of the main paper. We omit the moment subscript
$i$.  Conditional on a structural ensemble, let
$\boldsymbol\alpha=\delta\mathbf 1_K+\mathbf c$ and
$\alpha_0=\sum_k\alpha_k=M+K\delta$.  The multinomial probability of the
effective empirical count vector is
\begin{equation*}
 p(\mathbf n\mid\mathbf p)
 =\frac{\Gamma(N_{\mathrm{eff}}+1)}
        {\prod_{k=1}^K\Gamma(n_k+1)}
  \prod_{k=1}^Kp_k^{n_k}.
 \tag{A.1}
\end{equation*}
The structurally updated Dirichlet density is
\begin{equation*}
 p(\mathbf p\mid\Theta_{A,M})
 =\frac{\Gamma(\alpha_0)}{\prod_{k=1}^K\Gamma(\alpha_k)}
  \prod_{k=1}^Kp_k^{\alpha_k-1}.
 \tag{A.2}
\end{equation*}
Using the multivariate beta integral,
\begin{align*}
 p_{\mathrm{DM}}(\mathbf n\mid\Theta_{A,M})
 &=\int_{\Delta_{K-1}}
 p(\mathbf n\mid\mathbf p)
 p(\mathbf p\mid\Theta_{A,M})d\mathbf p \notag\\
 &=\frac{\Gamma(N_{\mathrm{eff}}+1)\Gamma(\alpha_0)}
        {\Gamma(N_{\mathrm{eff}}+\alpha_0)}
   \prod_{k=1}^K
   \frac{\Gamma(n_k+\alpha_k)}
        {\Gamma(n_k+1)\Gamma(\alpha_k)}.
 \tag{A.3}
\end{align*}
Substitution of $\alpha_k=c_k+\delta$ and
$\alpha_0=M+K\delta$ gives
equation~(8) of the main paper. No asymptotic approximation is involved.

For computation, equation~(A.3) is evaluated by summing log-gamma terms.
Underflow and overflow cells are treated like all other cells in the $K$-cell
partition and require no modification of the conjugate derivation. For every
included moment, the count vectors satisfy $\sum_kn_k=N_{\mathrm{eff}}$ and
$\sum_kc_k=M$.

\vspace{5mm}
\noindent
\textbf{Appendix B: Exchangeability, prior concentration, and boundary cases}

\vspace{3mm}
\textit{Exchangeable posterior.}  For any permutation $\sigma$ of the $M$
ensemble members, write
$\Theta_{A,M}^{\sigma}=(\theta_A^{\sigma(1)},\ldots,\theta_A^{\sigma(M)})$.
The structural cell counts satisfy
$c_{k,i}(\Theta_{A,M}^{\sigma})=c_{k,i}(\Theta_{A,M})$ for every $i$ and $k$.
Thus, the DMPI score $L_M$ and the product prior in
equation~(10) of the main paper are both invariant to $\sigma$, and so is the
generalized posterior $\Pi_{A,M}$.  For any measurable region $C$ of the
structural-parameter space and any $j,k\in\{1,\ldots,M\}$,
\begin{equation*}
 \Pi_{A,M}\{\theta_A^j\in C\}
 =\Pi_{A,M}\{\theta_A^k\in C\}
 =\mathbb E_{\Pi_{A,M}}[Q_M(C)].
\end{equation*}
The first equality follows by permuting the integration variables; the
second follows from the definition of $Q_M$ and linearity of expectation.
This common marginal is the DMPI posterior for a single structural-parameter
vector.

\textit{Product-prior concentration.}  Let $C_1,\ldots,C_L$ be a finite
measurable partition of the admissible parameter space and set
\begin{equation*}
 \bar\pi_l=\int_{C_l}\pi_A(\theta_A)\,d\theta_A,
 \qquad l=1,\ldots,L.
\end{equation*}
Under the product prior $\Pi_{A,M}^0$ in equation~(10) of the main paper, the
ensemble members are iid draws from $\pi_A$.  Hence the occupation vector
$\mathbf r$, with
$r_l=\sum_{j=1}^M\mathbf 1\{\theta_A^j\in C_l\}$, is multinomial with
probabilities $\bar{\boldsymbol\pi}$.  Let
$w_l=r_l/M=Q_M(C_l)$.
\begin{equation*}
 \Pr_{\Pi_{A,M}^0}(\mathbf r)
 =\frac{M!}{\prod_{l=1}^Lr_l!}
  \prod_{l=1}^L\bar\pi_l^{r_l}.
 \tag{B.1}
\end{equation*}
Applying Stirling's expansion to the multinomial coefficient gives
\begin{equation*}
 \log\Pr_{\Pi_{A,M}^0}(\mathbf r)
 =-M D_{\mathrm{KL}}(\mathbf w\Vert\bar{\boldsymbol\pi})
  +O(\log M).
 \tag{B.2}
\end{equation*}
Equation~(B.2) is exact at the multinomial level up to the displayed
Stirling remainder. It is the finite-partition form of the product-prior
concentration used in Section~2.5. The partition $\{C_l\}_{l=1}^L$ is only an
analytical device; the parameter space is not discretized in the DMPI
computation.

\textit{The $M=1$ endpoint.}  Suppress the moment subscript and suppose that
the single model-implied value falls in cell $h$.  Then $c_h=1$ and $c_k=0$
for $k\neq h$.  Relative to a factor constant in the structural parameter,
the exact DM score satisfies
\begin{equation*}
 p_{\mathrm{DM}}(\mathbf n\mid\Theta_{A,1})
 \propto n_h+\delta.
 \tag{B.3}
\end{equation*}
This follows from
$\Gamma(n_h+\delta+1)/\Gamma(\delta+1)
=(n_h+\delta)\Gamma(n_h+\delta)/\{\delta\Gamma(\delta)\}$.
Thus the score favors parameter values whose model-implied moment falls in a
cell receiving greater empirical mass, while $\delta$ prevents a cell with
zero empirical counts from receiving zero score.  If $N_{\mathrm{eff}}=1$
and $e$ denotes the cell in which the empirical count vector places its single
unit, then
\begin{equation*}
 p_{\mathrm{DM}}(\mathbf n\mid\Theta_{A,1})
 =\frac{\mathbf 1\{h=e\}+\delta}{1+K\delta}
 \longrightarrow \mathbf 1\{h=e\}
 \qquad\text{as }\delta\downarrow0.
 \tag{B.4}
\end{equation*}
This is the hard-tolerance ABC kernel associated with the chosen partition.

\textit{The MEI endpoint.}  Let $M\rightarrow\infty$ while
$N_{\mathrm{eff}}$, $K$, and $\delta$ remain fixed.  Couple the ensembles
across $M$ by taking $\Theta_{A,M}$ to contain the first $M$ members of an
infinite iid sequence from $\pi_A$.  The strong law of large numbers gives,
for every $k$ and $i$,
\begin{equation*}
 \frac{c_{k,i}(\Theta_{A,M})}{M}
 \xrightarrow{\mathrm{a.s.}} p_{A,k,i}
 \equiv
 \Pr_{\theta_A\sim\pi_A}\{g_{A,i}(\theta_A)\in B_{k,i}\}
 .
\end{equation*}
Assume that $p_{A,k,i}>0$ whenever $n_{k,i}>0$.  The exact score can be
written as
\begin{equation*}
 L_{i,M}(\Theta_{A,M})
 =\frac{N_{\mathrm{eff}}!}{\prod_{k=1}^K n_{k,i}!}
  \frac{\Gamma(M+K\delta)}
       {\Gamma(M+K\delta+N_{\mathrm{eff}})}
  \prod_{k=1}^K
  \frac{\Gamma(c_{k,i}+\delta+n_{k,i})}
       {\Gamma(c_{k,i}+\delta)}.
\end{equation*}
For fixed $a$ and $b$, the gamma-ratio expansion
$\Gamma(x+a)/\Gamma(x+b)\sim x^{a-b}$ gives
\begin{equation*}
 \frac{\Gamma(M+K\delta)}
      {\Gamma(M+K\delta+N_{\mathrm{eff}})}\sim M^{-N_{\mathrm{eff}}}.
\end{equation*}
For each cell with $n_{k,i}>0$, the same expansion gives
\begin{equation*}
 \frac{\Gamma(c_{k,i}+\delta+n_{k,i})}
      {\Gamma(c_{k,i}+\delta)}
 \sim c_{k,i}^{n_{k,i}}
 \sim(Mp_{A,k,i})^{n_{k,i}};
\end{equation*}
when $n_{k,i}=0$, this ratio equals one.  Since
$\sum_kn_{k,i}=N_{\mathrm{eff}}$,
\begin{equation*}
 M^{-N_{\mathrm{eff}}}\prod_{k=1}^K(Mp_{A,k,i})^{n_{k,i}}
 =M^{-N_{\mathrm{eff}}+\sum_kn_{k,i}}
  \prod_{k=1}^Kp_{A,k,i}^{n_{k,i}}
 =\prod_{k=1}^Kp_{A,k,i}^{n_{k,i}}.
\end{equation*}
It follows that
\begin{equation*}
 L_{i,M}(\Theta_{A,M})
 \xrightarrow{\mathrm{a.s.}}
 L_{A,i}^{\infty}
 \equiv
 \frac{N_{\mathrm{eff}}!}{\prod_{k=1}^K n_{k,i}!}
 \prod_{k=1}^Kp_{A,k,i}^{n_{k,i}}.
 \tag{B.5}
\end{equation*}
The right-hand side is the multinomial probability mass function evaluated at
the prior-predictive cell probabilities $\mathbf p_{A,i}$.  Equivalently, as
the total Dirichlet concentration $M+K\delta$ diverges, the conditional
Dirichlet distribution concentrates at $\mathbf p_{A,i}$, so its multinomial
mixture converges to an ordinary multinomial distribution.

\textit{Evidence and concentration of the score-relevant frequencies.} Each
moment-specific DM score is a probability mass and hence lies between
zero and one.  Because the limits in equation~(B.5) are deterministic, bounded
convergence applied to the shared product-prior ensemble gives
\begin{equation*}
 \lim_{M\rightarrow\infty}Z_{A,M}
 =\prod_{i=1}^I L_{A,i}^{\infty}.
 \tag{B.6}
\end{equation*}
Under the positivity condition above, this limit is strictly positive.

For $\varepsilon>0$, define the event that at least one cell frequency used
by the DMPI score lies outside an $\varepsilon$-neighborhood of its
prior-predictive probability:
\[
 \mathcal U_{M,\varepsilon}^{c}
 =
 \left\{
 \max_{1\leq i\leq I,\,1\leq k\leq K}
 \left|\frac{c_{k,i}(\Theta_{A,M})}{M}-p_{A,k,i}\right|
 \geq\varepsilon
 \right\}.
\]
Because $I$ and $K$ are fixed, the simultaneous strong law of large numbers
gives $\Pi_{A,M}^{0}(\mathcal U_{M,\varepsilon}^{c})\to0$. Since
$0\leq L_M\leq1$,
\begin{align*}
 \Pi_{A,M}(\mathcal U_{M,\varepsilon}^{c})
 &=\frac{
   \mathbb E_{\Pi_{A,M}^{0}}
   [L_M(\Theta_{A,M})\mathbf 1\{\mathcal U_{M,\varepsilon}^{c}\}]
  }{Z_{A,M}} \notag\\
 &\leq
 \frac{\Pi_{A,M}^{0}(\mathcal U_{M,\varepsilon}^{c})}
      {Z_{A,M}}
 \longrightarrow0.
 \tag{B.7}
\end{align*}
Thus, under the generalized posterior, every cell frequency entering the
declared DMPI score converges in probability to its prior-predictive
probability. Posterior reweighting therefore cannot alter the large-$M$ limit
relevant to the score, whereas the generalized evidence converges to the
nontrivial prior-predictive quantity in equation~(B.6). This establishes the
MEI endpoint stated in Section~2.5.

\vspace{5mm}
\noindent
\textbf{Appendix C: Annealed SMC normalizing constants and mutation kernels}

\vspace{3mm}
Define the unnormalized tempered measure and its normalizing constant by
\begin{equation*}
 \gamma_t(d\Theta)
 =L_M(\Theta)^{\beta_t}\Pi_{A,M}^0(d\Theta),
 \qquad
 Z_t=\int L_M(\Theta)^{\beta_t}\Pi_{A,M}^0(d\Theta),
\end{equation*}
so that $\Pi_{A,M,t}(d\Theta)=\gamma_t(d\Theta)/Z_t$.  Because
$\beta_0=0$ and the product prior is normalized, $Z_0=1$; because
$\beta_T=1$, $Z_T=Z_{A,M}$.

For two adjacent stages, write
\begin{equation*}
 L_M(\Theta)^{\beta_t}
 =L_M(\Theta)^{\beta_t-\beta_{t-1}}
  L_M(\Theta)^{\beta_{t-1}}.
\end{equation*}
Substitution into $Z_t$ gives
\begin{align*}
 Z_t
 &=\int
   L_M(\Theta)^{\beta_t-\beta_{t-1}}
   L_M(\Theta)^{\beta_{t-1}}
   \Pi_{A,M}^0(d\Theta).
\end{align*}
Dividing by $Z_{t-1}$ and using the definition of the normalized target at
stage $t-1$ yields
\begin{align*}
 \frac{Z_t}{Z_{t-1}}
 &=\int L_M(\Theta)^{\beta_t-\beta_{t-1}}
   \frac{L_M(\Theta)^{\beta_{t-1}}
         \Pi_{A,M}^0(d\Theta)}{Z_{t-1}} \notag\\
 &=\int L_M(\Theta)^{\beta_t-\beta_{t-1}}
   \Pi_{A,M,t-1}(d\Theta) \notag\\
 &=\mathbb E_{\Pi_{A,M,t-1}}
   \left[L_M(\Theta)^{\beta_t-\beta_{t-1}}\right].
 \tag{C.1}
\end{align*}
Thus the ratio $Z_t/Z_{t-1}$ is the expectation, under the previous tempered
target, of the incremental weight.  Equally weighted particles from
$\Pi_{A,M,t-1}$ estimate this ratio by
\begin{equation*}
 \widehat{\frac{Z_t}{Z_{t-1}}}
 =\frac{1}{P}\sum_{p=1}^P
   L_M(\Theta_{t-1}^p)^{\beta_t-\beta_{t-1}}
 =\frac{1}{P}\sum_{p=1}^P\widetilde w_t^p.
\end{equation*}
Finally, the normalizing constants telescope:
\begin{equation*}
 Z_{A,M}=Z_T
 =Z_0\prod_{t=1}^T\frac{Z_t}{Z_{t-1}}
 =\prod_{t=1}^T\frac{Z_t}{Z_{t-1}}.
\end{equation*}
Replacing each ratio by its particle estimate gives
equation~(13) of the main paper; taking logarithms turns the product into the
reported sum.  In the reported calculations, the schedule is fixed before the independent
runs used for evidence estimation.  Resampling changes the particle
representation but not the target or the identity in equation~(C.1).

Consider a draw-wise proposal that replaces column $j$ of $\Theta$ by
$\theta'$ drawn from $q_t(\theta'\mid\theta^j)$.  With $\Theta'$ denoting the
proposed ensemble, the Metropolis--Hastings probability is
\begin{equation*}
 a_t(\Theta,\Theta')
 =1\wedge
 \frac{L_M(\Theta')^{\beta_t}\pi_A(\theta')
       q_t(\theta^j\mid\theta')}
      {L_M(\Theta)^{\beta_t}\pi_A(\theta^j)
       q_t(\theta'\mid\theta^j)}.
 \tag{C.2}
\end{equation*}
Let $\pi_t$ denote the density of $\Pi_{A,M,t}$ and write
$q_t(\Theta'\mid\Theta)$ for the corresponding proposal density on the
ensemble space.  Define the accepted-move density
$r_t(\Theta'\mid\Theta)=q_t(\Theta'\mid\Theta)a_t(\Theta,\Theta')$.
Equation~(C.2) implies
\begin{align*}
 \pi_t(\Theta)r_t(\Theta'\mid\Theta)
 &=\pi_t(\Theta)q_t(\Theta'\mid\Theta)
   a_t(\Theta,\Theta') \\
 &=\pi_t(\Theta)q_t(\Theta'\mid\Theta)
   \left[1\wedge
   \frac{\pi_t(\Theta')q_t(\Theta\mid\Theta')}
        {\pi_t(\Theta)q_t(\Theta'\mid\Theta)}\right] \\
 &=\min\{\pi_t(\Theta)q_t(\Theta'\mid\Theta),
           \pi_t(\Theta')q_t(\Theta\mid\Theta')\} \\
 &=\pi_t(\Theta')q_t(\Theta\mid\Theta')
   a_t(\Theta',\Theta) \\
 &=\pi_t(\Theta')r_t(\Theta\mid\Theta').
\end{align*}
Define the rejection probability
\begin{equation*}
 \rho_t(\Theta)
 =1-\int r_t(u\mid\Theta)\,du.
\end{equation*}
Writing $\delta_{\Theta}(\Theta')$ for the Dirac point mass at $\Theta$, the
complete Metropolis--Hastings kernel has the generalized transition density
\begin{equation*}
 k_t(\Theta'\mid\Theta)
 =r_t(\Theta'\mid\Theta)
  +\rho_t(\Theta)\delta_{\Theta}(\Theta').
\end{equation*}
The accepted-move detailed balance above then gives
\begin{align*}
 \int \pi_t(\Theta)k_t(\Theta'\mid\Theta)\,d\Theta
 &=\int \pi_t(\Theta)r_t(\Theta'\mid\Theta)\,d\Theta
   +\pi_t(\Theta')\rho_t(\Theta') \\
 &=\pi_t(\Theta')
   \left\{\int r_t(\Theta\mid\Theta')\,d\Theta
   +\rho_t(\Theta')\right\} \\
 &=\pi_t(\Theta').
\end{align*}
Thus the complete kernel leaves $\Pi_{A,M,t}$ invariant.

The sampler uses two independence proposals.  For the prior-independence
move, $q_t=\pi_A$, so the prior terms in equation~(C.2) cancel exactly.  For
the fitted global move, let $z=T(\theta_A)$ be a one-to-one
differentiable transformation from the interior of the coordinate-wise
parameter support to $\mathbb R^d$.  At the beginning of annealing stage $t$,
a multivariate Student-$t$ density $\widetilde q_t(z)$ is fitted to the
transformed structural-parameter values represented by the particle
population.  A proposal is generated as
\begin{equation*}
 z'\sim\widetilde q_t,
 \qquad
 \theta_A'=T^{-1}(z').
\end{equation*}
The corresponding proposal density with respect to the original parameter
coordinates is
\begin{equation*}
 q_t(\theta_A')
 =\widetilde q_t\{T(\theta_A')\}
  \left|\det DT(\theta_A')\right|.
\end{equation*}
Because this is an independence proposal,
$q_t(\theta_A'\mid\theta_A^j)=q_t(\theta_A')$ and
$q_t(\theta_A^j\mid\theta_A')=q_t(\theta_A^j)$ in equation~(C.2).  The fitted
density is held fixed throughout the mutation steps of stage $t$ and may be
refitted between annealing stages.

Fixed-probability mixtures and compositions of invariant kernels also preserve
$\Pi_{A,M,t}$.  Hence local random-walk, prior-independence, and fitted global
moves may be combined.

\nobreak\vspace{2mm}
\noindent
\textbf{Appendix D: Correctly specified NKPC benchmark}

\vspace{1mm}
This appendix reports the counterpart to the RR Monte Carlo
experiment in Section~3.  The data-generating process, empirical VAR,
population moments, partitions, effective empirical counts, and numerical
settings are unchanged.  The structural model used for inference is now FR:
it includes both the output-gap shock $\epsilon_t$ and the NKPC shock $v_t$,
so it is correctly specified and has a full-rank innovation covariance matrix.
The additional structural parameter is $\sigma_v^2$. Its prior is an
underlying Normal distribution with location $0.00025$ and scale $0.0001$,
truncated below at zero. The priors for the four parameters shared with the
RR model are reported in Table~2 of the main paper.

\textit{Numerical implementation.} The reported calculations use $P=512$ particles and
three complete mutation sweeps at each annealing stage. After resampling, each
sweep visits every ensemble member in every particle and applies a
Metropolis--Hastings update conditional on the other $M-1$ members. The
model--$M$-specific temperature paths were constructed in preliminary calculations using a
target incremental-weight ESS of $0.5P$ and then held fixed. For
$\mathcal M=(1,10,50,100,200,300)$, the paths contain
$T=(4,20,41,55,75,89)$ stages for the RR model and
$T=(7,25,55,74,103,123)$ stages for the FR model. Prespecified verification
runs use an independent seed at $P=512$ and then double the particle count to
$P=1024$; fresh SMC seeds are used across the 30 simulated datasets.

\newpage
Table~\ref{tab:posterior_correct_v2} reports the posterior and evidence
results.  The posterior means of all five structural parameters remain close
to their DGP values over the full grid of $M$.  In particular, the posterior
mean of $\sigma_v^2$ remains close to $0.00025$, while the other four estimates
are similar to their RR counterparts.  As in the RR
experiment, increasing $M$ does not mechanically narrow the posterior
intervals because it strengthens structural discipline without increasing the
fixed empirical information scale $N_{\mathrm{eff}}$.

\renewcommand{\thetable}{D.\arabic{table}}
\setcounter{table}{0}
\begin{table}[H]
\centering
\captionsetup{font=footnotesize}
\caption{Monte Carlo results for the correctly specified FR NKPC model}
\label{tab:posterior_correct_v2}
\scriptsize
\setlength{\tabcolsep}{2.4pt}
\renewcommand{\arraystretch}{0.78}
\resizebox{\textwidth}{!}{%
\begin{tabular}{c|c|c|c|c|c|c}
\hline
$M$ & $\beta$ & $\mu_p$ & $\rho$ & $\sigma_\epsilon^2$ & $\sigma_v^2$ & $\log\widehat Z_{\mathrm{FR},M}$ \\
 & [0.980] & [0.800] & [0.800] & [0.001] & [0.00025] & \\
\hline\hline
1 & 0.97999 & 0.80177 & 0.79759 & 0.000999 & 0.000246 & -9119.58 \\
  & (0.97801, 0.98189) & (0.77425, 0.82843) & (0.74987, 0.84285) & (0.000886, 0.001120) & (0.000183, 0.000310) & (0.81) \\
\hline
10 & 0.97999 & 0.80168 & 0.79729 & 0.001002 & 0.000247 & -9006.94 \\
   & (0.97800, 0.98190) & (0.77064, 0.83157) & (0.74492, 0.84628) & (0.000879, 0.001133) & (0.000173, 0.000325) & (8.18) \\
\hline
50 & 0.98000 & 0.80152 & 0.79656 & 0.001004 & 0.000247 & -8675.58 \\
   & (0.97799, 0.98191) & (0.76607, 0.83544) & (0.73684, 0.85240) & (0.000862, 0.001155) & (0.000167, 0.000332) & (41.90) \\
\hline
100 & 0.98000 & 0.80146 & 0.79625 & 0.001004 & 0.000247 & -8424.33 \\
    & (0.97799, 0.98191) & (0.76459, 0.83678) & (0.73311, 0.85503) & (0.000855, 0.001166) & (0.000165, 0.000335) & (84.24) \\
\hline
200 & 0.98000 & 0.80145 & 0.79608 & 0.001005 & 0.000247 & -8142.88 \\
    & (0.97799, 0.98191) & (0.76346, 0.83776) & (0.73008, 0.85710) & (0.000850, 0.001174) & (0.000164, 0.000336) & (167.94) \\
\hline
300 & 0.98000 & 0.80145 & 0.79598 & 0.001005 & 0.000247 & -8005.67 \\
    & (0.97799, 0.98191) & (0.76294, 0.83828) & (0.72869, 0.85794) & (0.000848, 0.001177) & (0.000163, 0.000337) & (250.55) \\
\hline
\end{tabular}%
}
\vspace{1mm}
\captionsetup{font=scriptsize}
\caption*{\textit{Note.} Values in brackets are the DGP parameters. For each
structural parameter, the table reports the Monte Carlo average of the
posterior mean (top) and the Monte Carlo averages of the 2.5th and 97.5th
posterior percentiles (bottom). The final column reports the Monte Carlo mean
of the estimated log joint generalized evidence, with its Monte Carlo standard
deviation in parentheses. Results are based on the same 30 independently
simulated datasets used in Section~3.}
\end{table}

Figure~\ref{fig:correct_v2_feature_distributions} compares the empirical and
FR model-implied distributions.  All five population moments display
substantial overlap.  Some differences in dispersion remain, most visibly for
$a_{22}$, but there is no persistent displacement comparable to that of
$[\Sigma_e]_{11}$ under the RR model.  The inclusion of $v_t$ allows
the model to reproduce the innovation variance of the inflation equation
without forcing compensating changes in the other four moments.

\renewcommand{\thefigure}{D.\arabic{figure}}
\setcounter{figure}{0}
\begin{figure}[H]
    \centering
    \includegraphics[width=0.96\linewidth]{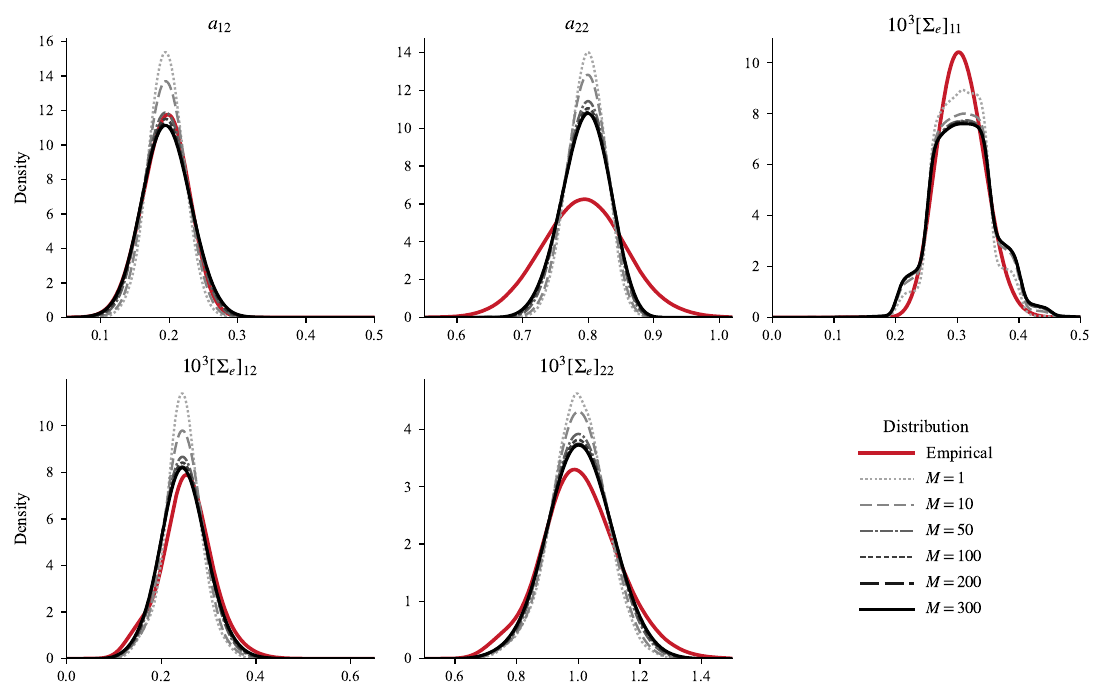}
    \captionsetup{font=footnotesize}
    \caption{Monte Carlo-averaged population-moment distributions for the
    correctly specified FR NKPC model}
    \label{fig:correct_v2_feature_distributions}
    \captionsetup{font=scriptsize}
    \caption*{\textit{Note.} The empirical curve pools the 50{,}000
    VAR-posterior population-moment draws from each of 30 simulated datasets. For each
    $M$, the model-implied curve pools the members of the final equally
    weighted SMC ensembles across the same datasets. All curves use the same
    smoothing bandwidths as Figure~1 of the main paper.
    The three covariance moments are multiplied by $10^3$.}
\end{figure}

The average log joint generalized evidence increases monotonically from
$-9119.58$ at $M=1$ to $-8005.67$ at $M=300$.  At the replication level,
$M=300$ maximizes the estimated evidence in 26 of the 30 datasets, and
$M=200$ does so in the remaining four.  Thus the correctly specified model
shows no systematic interior reversal over the prespecified grid.  This
contrasts with the RR model, whose evidence usually peaks at
$M=50$ and declines sharply thereafter.  The comparison supports the
interpretation of the RR model's hump-shaped profile as evidence of
tension between its omitted-shock restriction and the declared empirical
target, rather than as a generic consequence of increasing the ensemble size.

\vspace{2mm}
\noindent
\textit{Classical minimum-distance benchmark.}
We also compare the DMPI diagnosis with a conventional sample-moment-matching method
using the same 30 datasets. For each
dataset, let $\widehat{\mathbf m}$ collect the five sample moments used as the
CMD target: the OLS estimates of the quantities in
equation~(17) of the main paper from the unrestricted
Gaussian VAR(1). The residual covariance matrix is estimated by
the degree-of-freedom-adjusted OLS estimator, with residual degrees of freedom
$\nu=296$. Let $\widehat\Omega$ be the plug-in frequentist sampling covariance
of $\widehat{\mathbf m}$. Its first block is the standard Gaussian OLS
covariance of $(\widehat a_{12},\widehat a_{22})$, and its second is the
Wishart covariance of
$(\widehat\Sigma_{11},\widehat\Sigma_{12},\widehat\Sigma_{22})$. Under the
Gaussian VAR, the estimation errors in the coefficient and
innovation-covariance estimators are asymptotically uncorrelated, so their
cross-covariance is set to zero. CMD minimizes
\[
 J_{\mathrm{CMD}}(\theta_A)
 =\{b_A(\theta_A)-\widehat{\mathbf m}\}'\widehat\Omega^{-1}
  \{b_A(\theta_A)-\widehat{\mathbf m}\},
\]
where $b_A(\theta_A)$ is the analytical structural mapping into the five
moments. The calculation does not use the empirical-model posterior draws.
Because the five moments do not separately identify $\beta$, it is fixed at
its DGP value $0.98$ in both models. The FR and RR specifications then have
four and three estimated parameters, respectively.

Table~\ref{tab:nkpc_cmd_estimates} reports the corresponding parameter
estimates. The FR estimates are centered close to the DGP values, and the RR
estimates of $\mu_p$ and $\rho$ also remain close to their DGP values.

\begin{table}[H]
\centering
\captionsetup{font=footnotesize}
\caption{Classical minimum-distance parameter estimates}
\label{tab:nkpc_cmd_estimates}
\scriptsize
\setlength{\tabcolsep}{6pt}
\renewcommand{\arraystretch}{0.92}
\begin{tabular}{lrrrr}
\hline
Model & $\mu_p$ & $\rho$ & $10^3\sigma_\epsilon^2$
& $10^3\sigma_v^2$ \\
\hline
DGP & 0.800 & 0.800 & 1.000 & 0.250 \\
\hline
FR & 0.800 & 0.794 & 1.007 & 0.243 \\
   & (0.023) & (0.045) & (0.098) & (0.022) \\
RR & 0.800 & 0.793 & 1.004 & --- \\
   & (0.024) & (0.045) & (0.097) &  \\
\hline
\end{tabular}
\par\vspace{1mm}
\begin{minipage}{0.88\textwidth}
\scriptsize
\textit{Note.} Entries are Monte Carlo means across 30 replications, with
Monte Carlo standard deviations in parentheses.  The discount factor is fixed
at its DGP value, $\beta=0.98$.  The RR model imposes $\sigma_v^2=0$ rather
than estimating it.
\end{minipage}
\end{table}

Table~\ref{tab:nkpc_cmd_benchmark} reports the minimized CMD criterion and its
comparison with the conventional reference values.

\begin{table}[H]
\centering
\captionsetup{font=footnotesize}
\caption{Classical minimum-distance benchmark}
\label{tab:nkpc_cmd_benchmark}
\scriptsize
\setlength{\tabcolsep}{7pt}
\renewcommand{\arraystretch}{0.90}
\begin{tabular}{lrr}
\hline
Model & Mean minimized criterion & Fraction above reference \\
\hline
FR & 0.914 & 0.033 \\
RR & 148.000 & 1.000 \\
\hline
\end{tabular}
\par\vspace{1mm}
\begin{minipage}{0.88\textwidth}
\scriptsize
\textit{Note.} Entries average 30 Monte Carlo replications. The last column
uses the conventional 5\% upper-tail reference values of $\chi^2_1$ for FR
and $\chi^2_2$ for RR. For RR this fraction is descriptive rather than a
formal rejection rate: stochastic singularity violates the nonsingularity
conditions underlying the regular chi-square reference. The calculations use
the immutable dataset from each corresponding DMPI replication, but not its
empirical-model posterior draws.
\end{minipage}
\end{table}

The CMD criterion exceeds its conventional reference value for the correctly
specified FR model in one of the 30 replications and for the RR model in every
replication. The mean difference $J_{\mathrm{RR}}-J_{\mathrm{FR}}$ is
147.086. Thus, although the RR parameter estimates in
Table~\ref{tab:nkpc_cmd_estimates} are close to their DGP values, the
best-fitting RR parameter values still leave a large joint discrepancy across
the five sample moments. The same omitted-shock tension is reflected in the
DMPI $M$ profile.

We do not report SMM using the same unrestricted VAR auxiliary model. Under RR,
$\Delta\pi_t=b\phi_t$ holds exactly, making the simulated VAR regressors
perfectly collinear and their unrestricted coefficient estimator nonunique.
SMM based on that estimator would depend on arbitrary regularization, while
indirect inference using it would have a nonunique binding function.

\newpage
\noindent
\textbf{Appendix E: Additional results and sensitivity checks for the U.S. application}

\vspace{3mm}
\noindent
\textit{Posterior distributions for the RR specifications.}
Tables~\ref{tab:posterior_rr_data}--\ref{tab:posterior_rrR_data}
report posterior summaries for the three RR specifications compared
in Section~4.4.2. Figures~\ref{fig:rrg_parameter_appendix}--
\ref{fig:rrR_moments_appendix} show their structural-parameter and
population-moment distributions. The RR\textsubscript{g} figures use
$M\in\{1,2,5,10,20,30,50\}$, whereas the RR\textsubscript{z} and RR\textsubscript{R} figures use the
common comparison values $M\in\{1,2,10,50\}$. These figures provide
distributional detail for the joint generalized-evidence comparison reported
in Table~6 of the main paper. In the parameter figures, solid curves are posterior marginal
densities and the dashed curve is the conditional base-prior marginal. In the
population-moment figures, solid curves are model-implied distributions and
the dashed curve is the empirical distribution. The plotted distributions
combine the two independent SMC runs with equal weight. For the
population-moment figures, the same number of structural-ensemble members is
drawn from each run.

\renewcommand{\thetable}{E.\arabic{table}}
\setcounter{table}{0}
\input{appendix_rr_posterior_tables.tex}

\clearpage

\renewcommand{\thefigure}{E.\arabic{figure}}
\setcounter{figure}{0}
\begingroup
\setlength{\floatsep}{6pt}

\begin{figure}[p]
 \centering
 \includegraphics[width=0.82\linewidth]{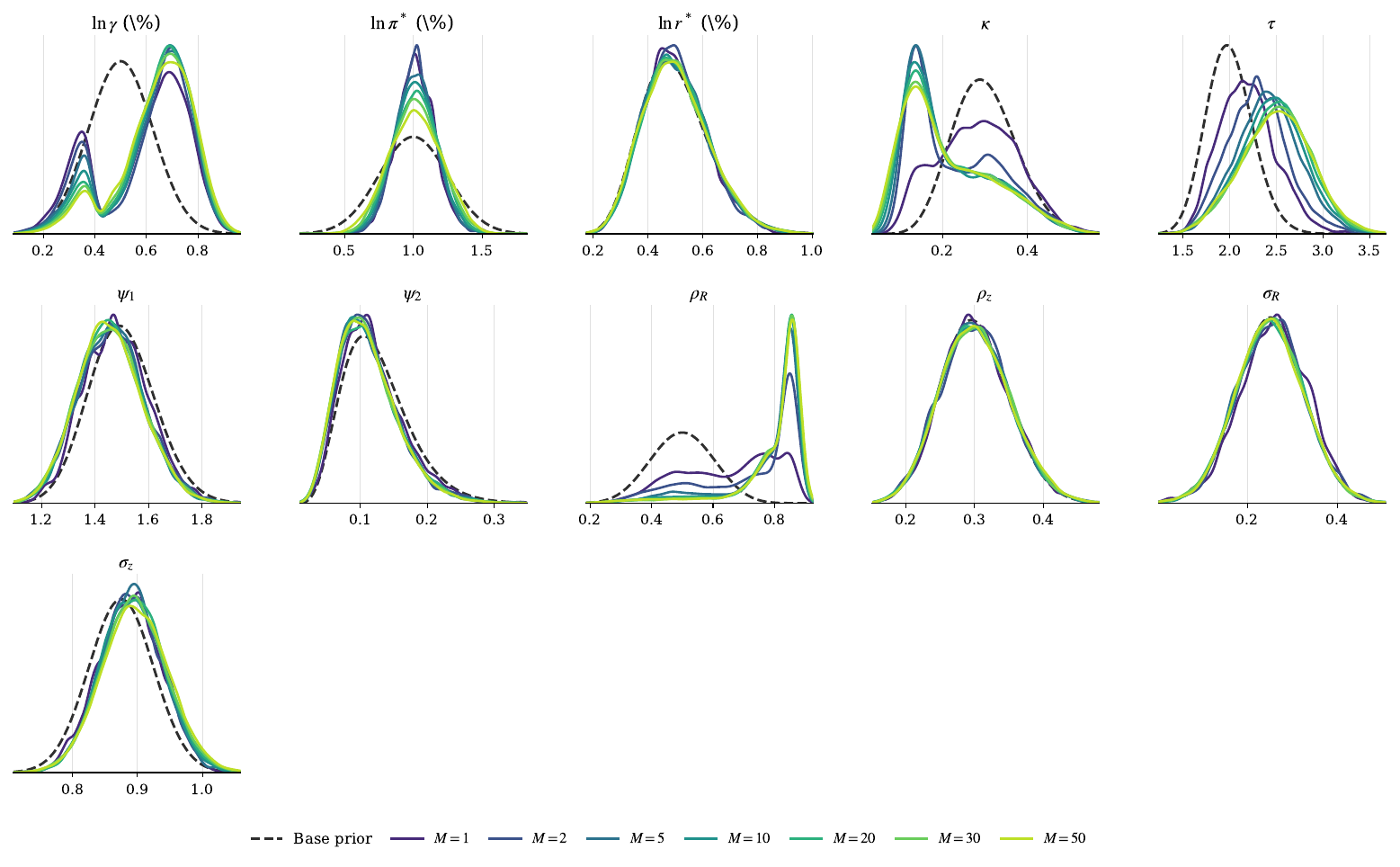}
 \captionsetup{font=footnotesize}
 \caption{Structural-parameter distributions for RR\textsubscript{g}}
 \label{fig:rrg_parameter_appendix}
\end{figure}

\begin{figure}[p]
 \centering
 \includegraphics[width=0.82\linewidth]{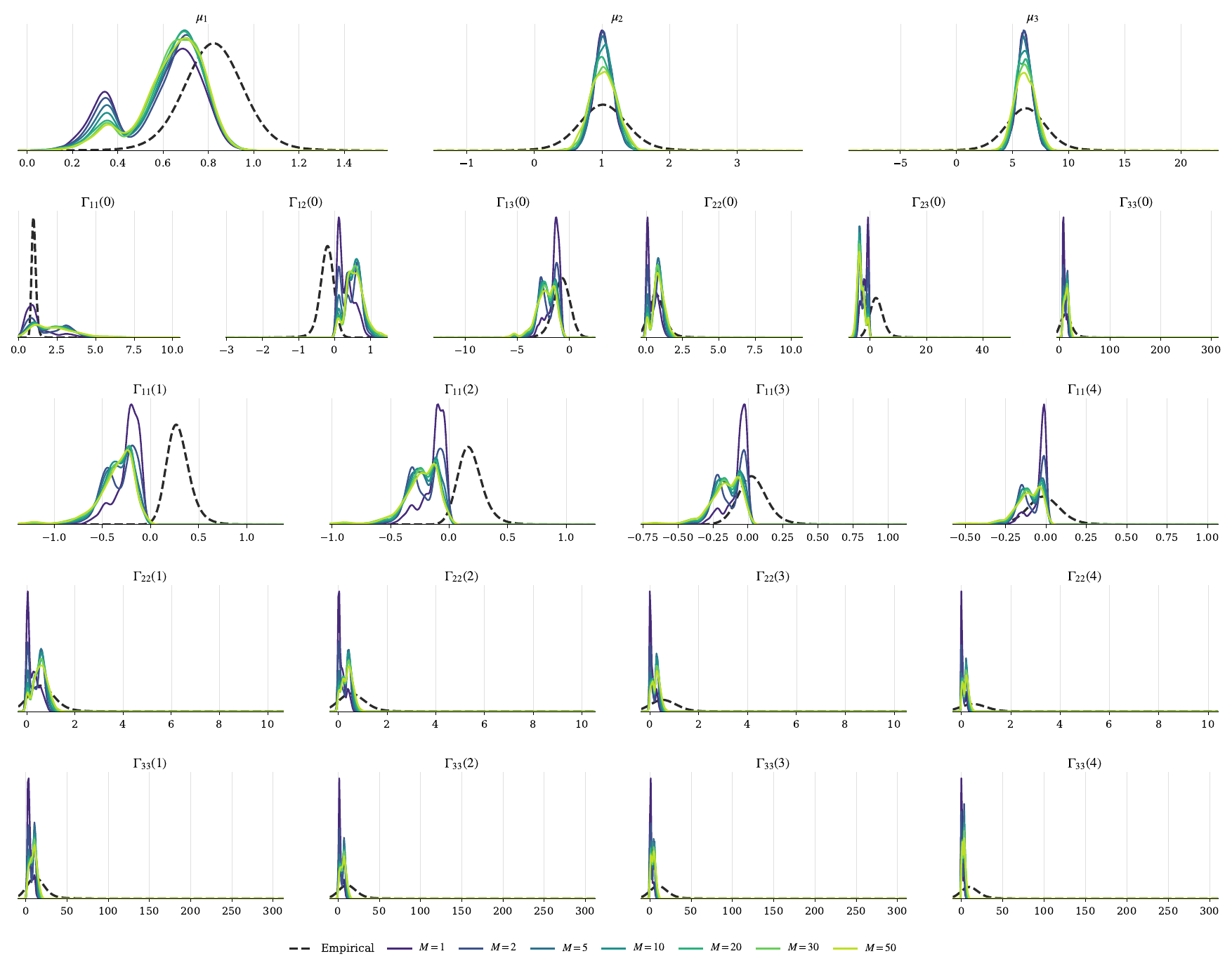}
 \captionsetup{font=footnotesize}
 \caption{Population-moment distributions for RR\textsubscript{g}}
 \label{fig:rrg_moments_appendix}
\end{figure}

\begin{figure}[p]
 \centering
 \includegraphics[width=0.82\linewidth]{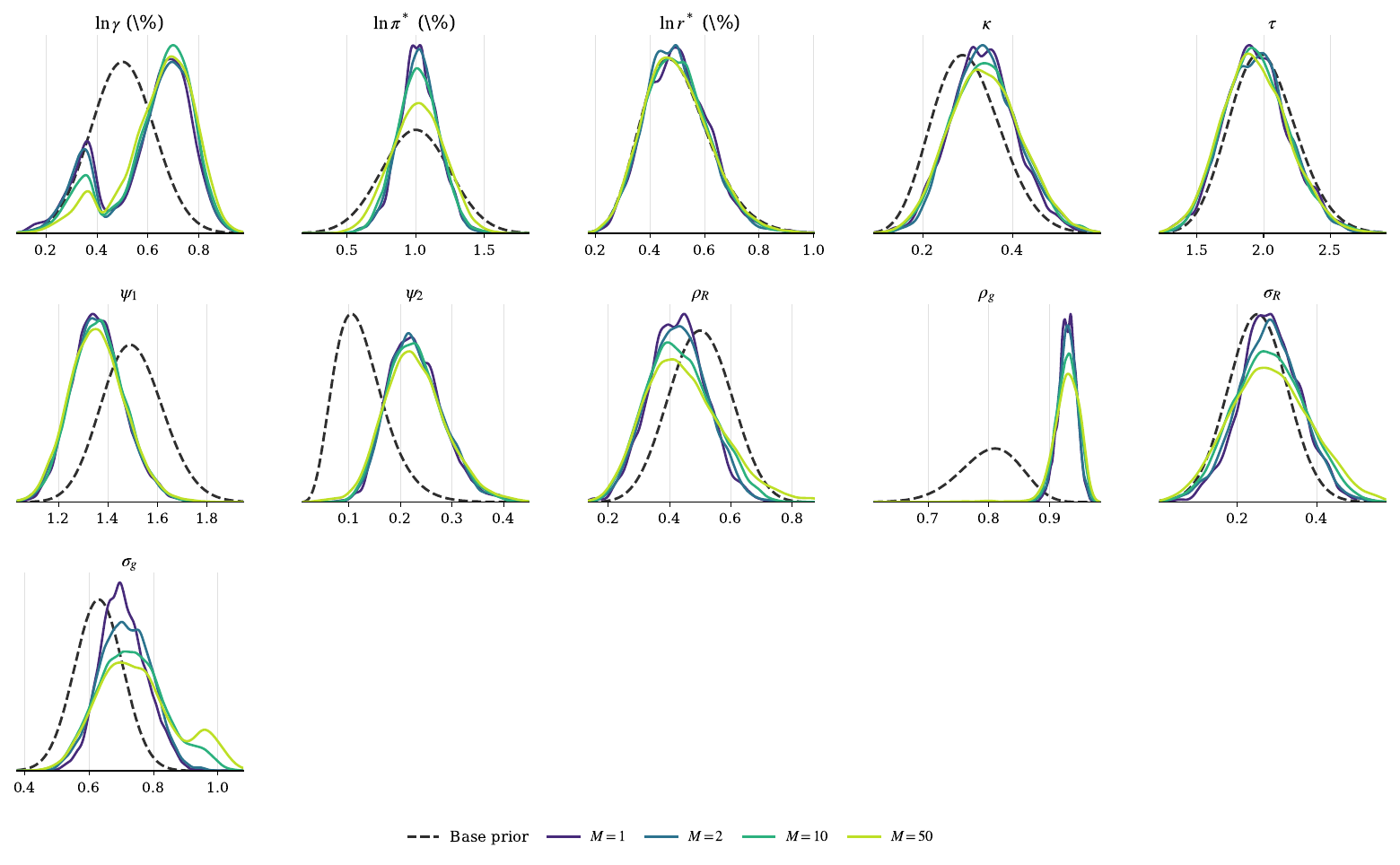}
 \captionsetup{font=footnotesize}
 \caption{Structural-parameter distributions for RR\textsubscript{z}}
 \label{fig:rrz_parameter_appendix}
\end{figure}

\begin{figure}[p]
 \centering
 \includegraphics[width=0.82\linewidth]{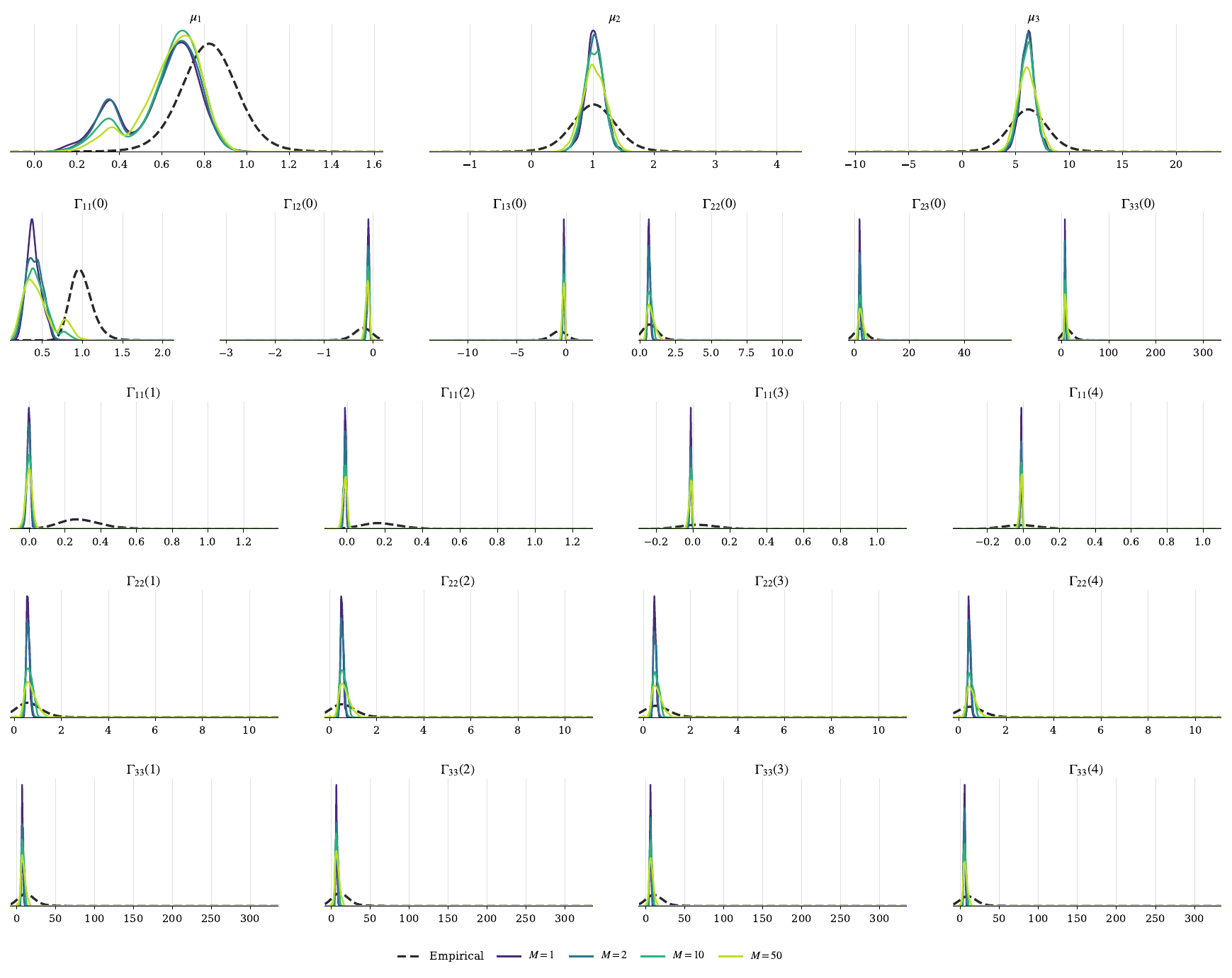}
 \captionsetup{font=footnotesize}
 \caption{Population-moment distributions for RR\textsubscript{z}}
 \label{fig:rrz_moments_appendix}
\end{figure}

\begin{figure}[p]
 \centering
 \includegraphics[width=0.82\linewidth]{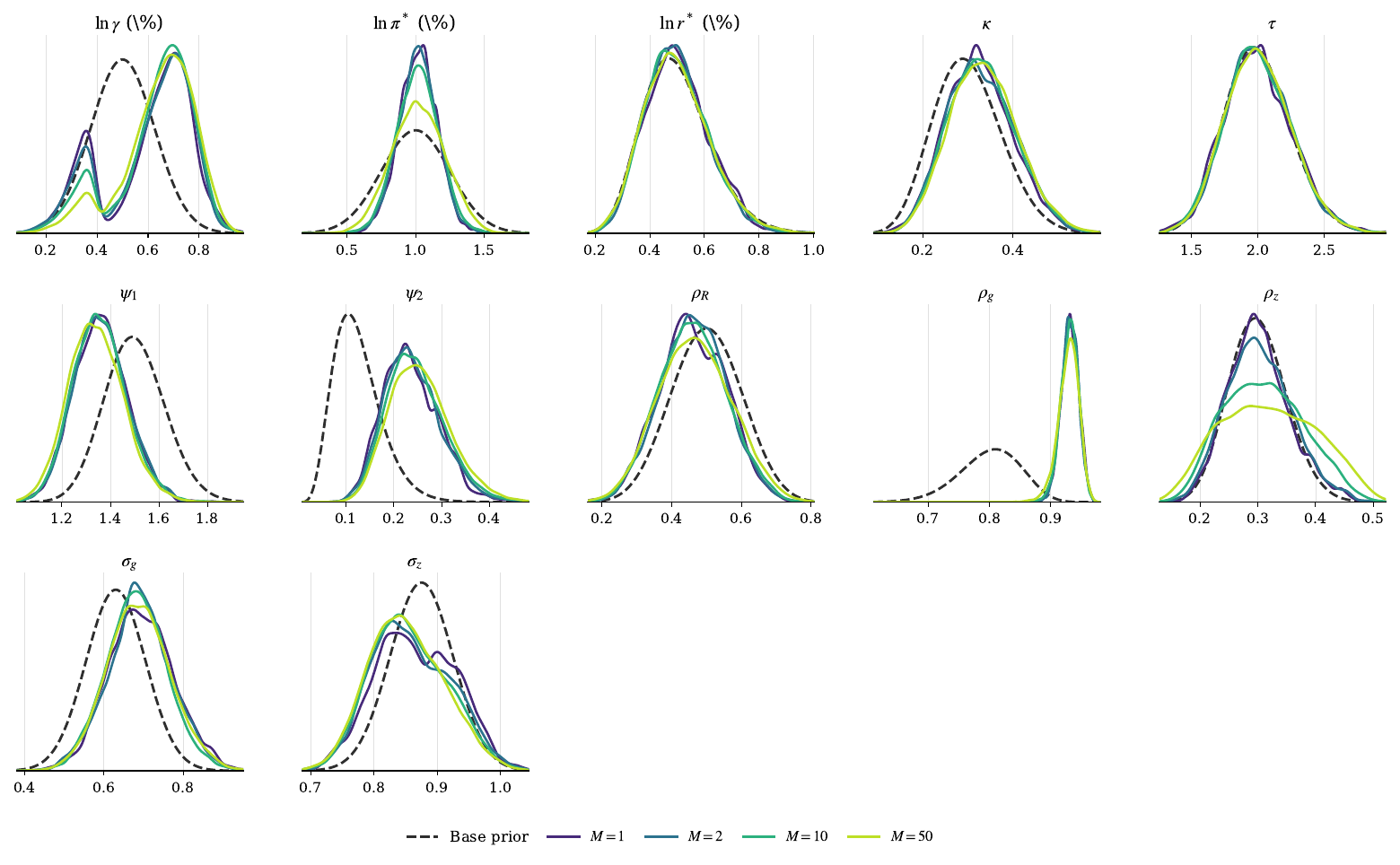}
 \captionsetup{font=footnotesize}
 \caption{Structural-parameter distributions for RR\textsubscript{R}}
 \label{fig:rrR_parameter_appendix}
\end{figure}

\begin{figure}[p]
 \centering
 \includegraphics[width=0.82\linewidth]{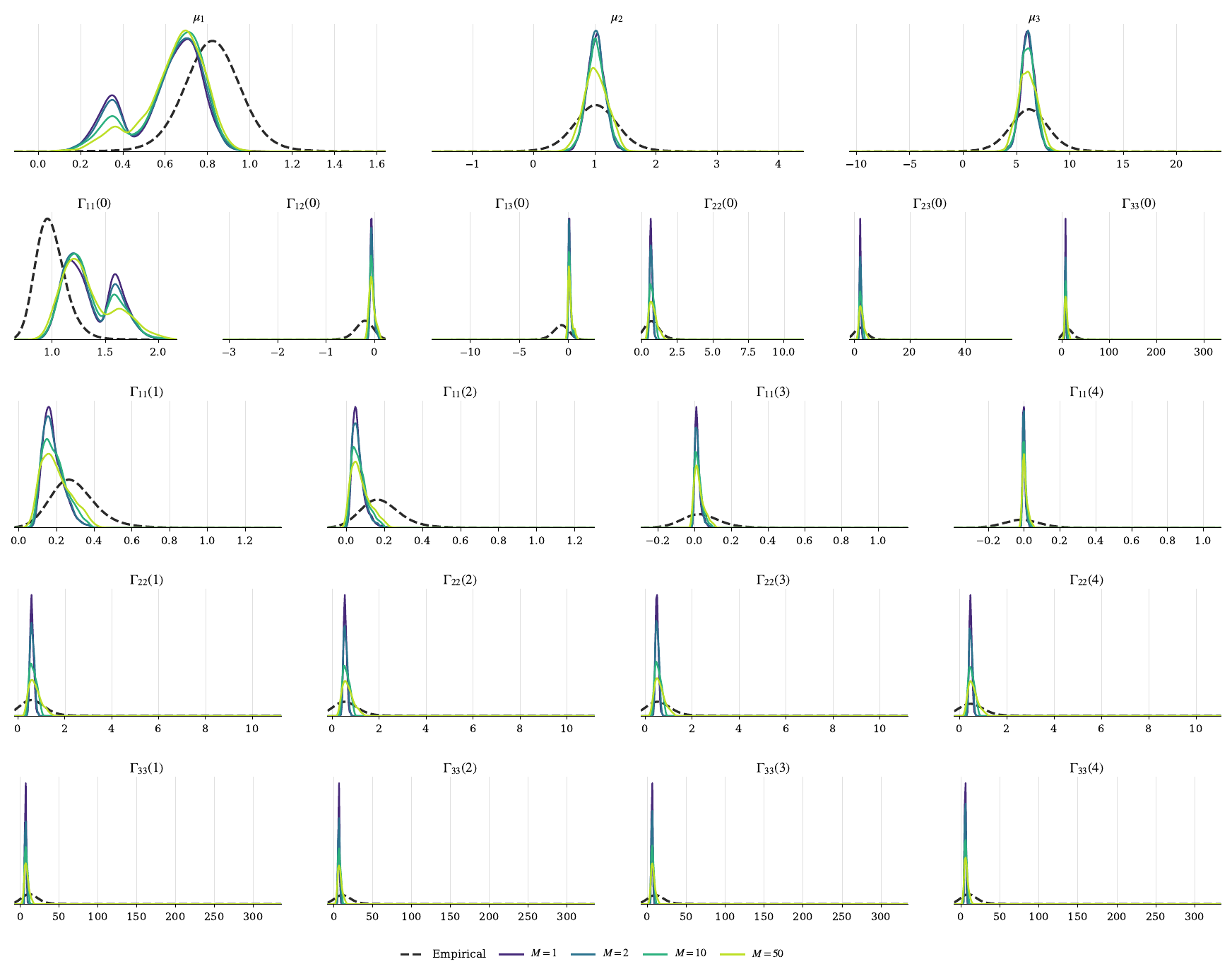}
 \captionsetup{font=footnotesize}
 \caption{Population-moment distributions for RR\textsubscript{R}}
 \label{fig:rrR_moments_appendix}
\end{figure}

\clearpage
\endgroup

\vspace{2mm}
\noindent
\textit{Numerical verification of the reported SMC calculations.}
All reported U.S. calculations use
$N_{\mathrm{MC}}=N_{\mathrm{eff}}=50{,}000$. For each model--$M$ cell, the
reported log generalized evidence is the log-mean-exp of the independent
estimates at the larger particle count. Table~\ref{tab:a10_numerical_diagnostics}
summarizes the numerical diagnostics across the $M$ grid for each model. All
22 cells reached $\beta=1$ without numerical failures.

\begin{table}[H]
\centering
\captionsetup{font=footnotesize}
\caption{Numerical verification of the baseline U.S. SMC estimates}
\label{tab:a10_numerical_diagnostics}
\scriptsize
\setlength{\tabcolsep}{4.0pt}
\renewcommand{\arraystretch}{0.90}
\begin{adjustbox}{max width=\textwidth}
\begin{tabular}{lclrrrr}
\hline
Model & $M$ grid & Reported $P$ & SD$(\log\widehat Z)$
& $|\Delta_P|$ & Min. ESS$/P$ & Max. wt. (\%) \\
\hline
FR & $1,2,5,10,20,30,50$ & $4096$--$8192$
& $0.130$--$0.356$ & $0.009$--$0.339$ & $0.790$ & $0.38$ \\
RR\textsubscript{g} & $1,2,5,10,20,30,50$ & $1024$
& $0.019$--$0.288$ & $0.002$--$0.188$ & $0.773$ & $1.43$ \\
RR\textsubscript{z} & $1,2,10,50$ & $1024$
& $0.073$--$0.261$ & $0.046$--$0.264$ & $0.700$ & $1.64$ \\
RR\textsubscript{R} & $1,2,10,50$ & $1024$
& $0.087$--$0.392$ & $0.028$--$0.335$ & $0.747$ & $1.72$ \\
\hline
\end{tabular}
\end{adjustbox}
\par\vspace{1mm}
\begin{minipage}{0.96\textwidth}
\scriptsize
\textit{Note.} Ranges are taken across the reported $M$ values for each model.
``Reported $P$'' is the larger particle count used for the reported evidence.
SD$(\log\widehat Z)$ is the between-run standard deviation at the smaller
particle count, and $|\Delta_P|$ is the absolute change in the log-mean-exp
estimate at the last particle doubling. ``Min. ESS$/P$'' and ``Max. wt.'' are,
respectively, the minimum incremental-weight effective-sample-size ratio and
the maximum normalized incremental weight across runs at both particle counts.
\end{minipage}
\end{table}

Across all cells, the largest between-run standard deviation at the smaller
particle count is $0.392$, and the largest absolute change at the last particle
doubling is $0.339$. These diagnostics support the numerical stability of the
reported generalized-evidence profiles.

\vspace{2mm}
\noindent
\textit{Robustness of the FR--RR\textsubscript{R} evidence comparison.}
We examine whether the ranking of RR\textsubscript{R} above FR in
Table~6 of the main paper depends on three features of the baseline
analysis.  The first check removes the lag-three and lag-four own
autocovariances, leaving 15 of the 21 prespecified population moments.  The
second rebuilds the empirical target from a VAR(2), rather than the baseline
VAR(4), while retaining $N_{\mathrm{MC}}=N_{\mathrm{eff}}=50{,}000$ and all
other target conventions.  The third changes only the location parameter of
the underlying Normal prior for the FR monetary-policy shock scale,
$s_R\sim\mathcal N(\mu_R,0.075^2)$ with $\sigma_R=|s_R|$, from the baseline
$\mu_R=0.251$ to either $0.1255$ or zero.  The corresponding
RR\textsubscript{R} results are unchanged in this third check because that model
does not contain $\sigma_R$.

\begin{table}[h]
\centering
\captionsetup{font=footnotesize}
\caption{Robustness of the FR--RR\textsubscript{R} generalized-evidence difference}
\label{tab:rrR_robust_evidence}
\scriptsize
\setlength{\tabcolsep}{2.7pt}
\renewcommand{\arraystretch}{0.83}
\begin{adjustbox}{max width=\textwidth}
\begin{tabular}{llrrrrrr}
\hline
Specification & $M$ & $\log\widehat Z_{\mathrm{FR},M}$
& $\log\widehat Z_{\mathrm{RR}_{R},M}$ & $\widehat\Delta_M$
& Baseline $\widehat\Delta_M$ & Change & MC SD \\
\hline
Baseline & 1  & -15204.89 & -15201.88 &  -3.02 &  -3.02 &   0.00 & 0.19 \\
         & 2  & -15196.59 & -15189.45 &  -7.14 &  -7.14 &   0.00 & 0.22 \\
         & 10 & -15185.43 & -15131.00 & -54.43 & -54.43 &   0.00 & 0.13 \\
         & 50 & -15639.22 & -15311.87 & -327.35 & -327.35 & 0.00 & 0.05 \\
\hline
15 moments & 1  & -10867.83 & -10865.33 &  -2.50 &  -3.02 &  0.51 & 0.26 \\
           & 2  & -10869.38 & -10863.20 &  -6.18 &  -7.14 &  0.96 & 0.28 \\
           & 10 & -10917.44 & -10870.13 & -47.31 & -54.43 &  7.13 & 0.56 \\
           & 50 & -11506.46 & -11201.62 & -304.84 & -327.35 & 22.51 & 0.33 \\
\hline
VAR(2) target & 1  & -15213.28 & -15208.01 &  -5.27 &  -3.02 &  -2.26 & 0.51 \\
              & 2  & -15212.79 & -15202.31 & -10.49 &  -7.14 &  -3.35 & 0.32 \\
              & 10 & -15245.03 & -15187.44 & -57.59 & -54.43 &  -3.15 & 0.13 \\
              & 50 & -15800.67 & -15496.42 & -304.25 & -327.35 & 23.10 & 0.80 \\
\hline
$\mu_R=0.1255$ & 1  & -15207.61 & -15201.88 &  -5.73 &  -3.02 &  -2.71 & 0.48 \\
                 & 2  & -15203.05 & -15189.45 & -13.60 &  -7.14 &  -6.46 & 0.11 \\
                 & 10 & -15216.55 & -15131.00 & -85.55 & -54.43 & -31.11 & 2.39 \\
                 & 50 & -15788.67 & -15311.87 & -476.79 & -327.35 & -149.44 & 0.61 \\
\hline
$\mu_R=0$ & 1  & -15213.08 & -15201.88 & -11.20 &  -3.02 &  -8.19 & 1.96 \\
           & 2  & -15214.00 & -15189.45 & -24.55 &  -7.14 & -17.41 & 1.55 \\
           & 10 & -15253.89 & -15131.00 & -122.89 & -54.43 & -68.46 & 5.59 \\
           & 50 & -15965.34 & -15311.87 & -653.47 & -327.35 & -326.12 & 1.78 \\
\hline
\end{tabular}%
\end{adjustbox}
\par\vspace{1mm}
\begin{minipage}{0.96\textwidth}
\scriptsize
\textit{Note.} Negative values of $\widehat\Delta_M$ favor
RR\textsubscript{R}. ``Change'' is the robustness value minus its baseline
counterpart. MC SD is the root-sum-square of the two models' across-seed
standard deviations at the larger particle count. The baseline and prior
checks reuse the registered RR\textsubscript{R} cohort because changing the
FR prior for $\sigma_R$ cannot affect a model in which
$\sigma_R$ is absent. Absolute log evidence under the 15-moment target is not
comparable with that under a 21-moment target.
\end{minipage}
\end{table}

Table~\ref{tab:rrR_robust_evidence} reports
$\widehat\Delta_M=\log\widehat Z_{\mathrm{FR},M}
-\log\widehat Z_{\mathrm{RR}_{R},M}$, so a negative value favors
RR\textsubscript{R}.  The difference is negative for every value of $M$ under
every check.  Removing the six longer-lag moments narrows the gap but preserves
its sign, with the smallest absolute difference equal to 2.50 at $M=1$.
Changing the empirical VAR order widens the gap for $M\in\{1,2,10\}$ and
narrows it for $M=50$, again without changing the ranking.  Moving the
$\sigma_R$ prior location toward zero widens the evidence gap at every $M$.
Because the 15-moment specification contains fewer moment-specific score
components, its absolute evidence levels are not compared with those under the
21-moment specifications; only the within-specification FR-
RR\textsubscript{R} difference is relevant.

Table~\ref{tab:rrR_sigma_prior} reports the corresponding FR
posterior summaries for $\sigma_R=|s_R|$.  Under the zero-location prior, the
posterior median remains between 0.215 and 0.224 across the four values of
$M$.  The posterior probability below 0.05 ranges from 0.026 to 0.071, while
the probability below 0.10 ranges from 0.047 to 0.120.  Thus the posterior
does not concentrate at zero even when the prior is centered there.  This
finding should not be read as excluding zero-neighborhood values altogether;
rather, allowing substantially more prior mass near zero does not improve the
FR evidence or overturn the RR\textsubscript{R} ranking.

\begin{table}[H]
\centering
\captionsetup{font=footnotesize}
\caption{Posterior sensitivity of the FR monetary-policy shock scale}
\label{tab:rrR_sigma_prior}
\scriptsize
\setlength{\tabcolsep}{2.4pt}
\renewcommand{\arraystretch}{0.82}
\begin{adjustbox}{max width=\textwidth}
\begin{tabular}{llrrrrrrr}
\hline
Specification & $M$ & Mean & Median & 90\% interval & 95\% interval
& $\Pr(\sigma_R<0.025)$ & $\Pr(\sigma_R<0.05)$ & $\Pr(\sigma_R<0.10)$ \\
\hline
15 moments & 1  & 0.322 & 0.331 & [0.160, 0.457] & [0.134, 0.478] & 0.001 & 0.002 & 0.010 \\
           & 2  & 0.329 & 0.338 & [0.168, 0.463] & [0.137, 0.484] & 0.000 & 0.001 & 0.008 \\
           & 10 & 0.345 & 0.354 & [0.189, 0.468] & [0.156, 0.489] & 0.000 & 0.001 & 0.005 \\
           & 50 & 0.350 & 0.358 & [0.198, 0.470] & [0.163, 0.492] & 0.000 & 0.001 & 0.005 \\
\hline
VAR(2) target & 1  & 0.359 & 0.359 & [0.264, 0.458] & [0.246, 0.476] & 0.000 & 0.000 & 0.000 \\
              & 2  & 0.360 & 0.359 & [0.264, 0.459] & [0.245, 0.479] & 0.000 & 0.000 & 0.000 \\
              & 10 & 0.357 & 0.358 & [0.251, 0.459] & [0.224, 0.479] & 0.000 & 0.000 & 0.001 \\
              & 50 & 0.358 & 0.358 & [0.251, 0.463] & [0.223, 0.484] & 0.000 & 0.000 & 0.001 \\
\hline
$\mu_R=0.1255$ & 1  & 0.293 & 0.292 & [0.201, 0.387] & [0.184, 0.405] & 0.000 & 0.000 & 0.000 \\
                 & 2  & 0.293 & 0.291 & [0.202, 0.388] & [0.184, 0.407] & 0.000 & 0.000 & 0.000 \\
                 & 10 & 0.280 & 0.286 & [0.149, 0.384] & [0.092, 0.404] & 0.005 & 0.011 & 0.028 \\
                 & 50 & 0.278 & 0.284 & [0.141, 0.385] & [0.086, 0.405] & 0.006 & 0.012 & 0.031 \\
\hline
$\mu_R=0$ & 1  & 0.211 & 0.219 & [0.043, 0.315] & [0.021, 0.335] & 0.030 & 0.056 & 0.095 \\
           & 2  & 0.221 & 0.224 & [0.105, 0.320] & [0.048, 0.338] & 0.013 & 0.026 & 0.047 \\
           & 10 & 0.206 & 0.215 & [0.041, 0.310] & [0.019, 0.328] & 0.032 & 0.059 & 0.102 \\
           & 50 & 0.203 & 0.215 & [0.033, 0.310] & [0.016, 0.328] & 0.038 & 0.071 & 0.120 \\
\hline
Baseline & 1  & 0.371 & 0.370 & [0.272, 0.473] & [0.255, 0.494] & 0.000 & 0.000 & 0.000 \\
         & 2  & 0.369 & 0.368 & [0.271, 0.471] & [0.253, 0.492] & 0.000 & 0.000 & 0.000 \\
         & 10 & 0.368 & 0.367 & [0.267, 0.471] & [0.248, 0.492] & 0.000 & 0.000 & 0.000 \\
         & 50 & 0.364 & 0.365 & [0.254, 0.472] & [0.227, 0.494] & 0.000 & 0.000 & 0.001 \\
\hline
\end{tabular}%
\end{adjustbox}
\par\vspace{1mm}
\begin{minipage}{0.96\textwidth}
\scriptsize
\textit{Note.} Entries pool the final, equally weighted SMC particles from
the two independent runs at the reported particle count. Intervals are
equal-tailed posterior intervals. The probabilities refer to
$\sigma_R=|s_R|$. The baseline prior has $\mu_R=0.251$ and scale 0.075; the
two prior checks change only $\mu_R$.
\end{minipage}
\end{table}

\newpage
\vspace{2mm}
\noindent
\textit{Additional design robustness of the FR--RR\textsubscript{R}
evidence comparison.}
We examine three further prespecified changes to the empirical-target and
prior design.  The first expands the interior partition range from the
baseline empirical quantiles $(0.005,0.995)$ to $(0.001,0.999)$ while retaining
$K=102$.  The second increases the number of cells from $K=102$ to $K=202$
while retaining the baseline interior range.  Because $\delta=1$ is held fixed,
this change also raises the total Dirichlet base mass $K\delta$ from 102 to
202; it therefore does not isolate partition resolution.  The third doubles
the widths of the folded-Normal shock priors while leaving their locations and
all non-shock priors unchanged.

\begin{table}[H]
\centering
\captionsetup{font=footnotesize}
\caption{Additional design robustness of the FR--RR\textsubscript{R}
generalized-evidence difference}
\label{tab:rrR_design_robust_evidence}
\scriptsize
\setlength{\tabcolsep}{2.7pt}
\renewcommand{\arraystretch}{0.83}
\begin{adjustbox}{max width=\textwidth}
\begin{tabular}{llrrrrrr}
\hline
Specification & $M$ & $\log\widehat Z_{\mathrm{FR},M}$
& $\log\widehat Z_{\mathrm{RR}_{R},M}$ & $\widehat\Delta_M$
& Baseline $\widehat\Delta_M$ & Change & MC SD \\
\hline
D1: 0.1--99.9\% range ($K=102$)
  & 1  & -15190.76 & -15181.60 &  -9.17 &  -3.02 &  -6.15 & 0.10 \\
  & 2  & -15167.16 & -15150.80 & -16.36 &  -7.14 &  -9.22 & 0.29 \\
  & 10 & -15078.53 & -14995.70 & -82.83 & -54.43 & -28.39 & 0.25 \\
  & 50 & -15360.57 & -14940.06 & -420.52 & -327.35 & -93.16 & 0.05 \\
\hline
D2: $K=202$ (0.5--99.5\% range)
  & 1  & -27427.33 & -27425.34 &  -1.99 &  -3.02 &   1.02 & 0.13 \\
  & 2  & -27417.04 & -27412.01 &  -5.03 &  -7.14 &   2.11 & 0.09 \\
  & 10 & -27374.47 & -27331.10 & -43.38 & -54.43 &  11.06 & 0.18 \\
  & 50 & -27589.43 & -27294.66 & -294.77 & -327.35 & 32.59 & 0.36 \\
\hline
D3: doubled shock-prior widths
  & 1  & -15203.07 & -15200.57 &  -2.50 &  -3.02 &  0.52 & 0.32 \\
  & 2  & -15192.74 & -15186.92 &  -5.82 &  -7.14 &  1.32 & 0.25 \\
  & 10 & -15159.98 & -15119.39 & -40.59 & -54.43 & 13.84 & 0.21 \\
  & 50 & -15487.58 & -15244.88 & -242.70 & -327.35 & 84.66 & 0.29 \\
\hline
\end{tabular}%
\end{adjustbox}
\par\vspace{1mm}
\begin{minipage}{0.96\textwidth}
\scriptsize
\textit{Note.} $\widehat\Delta_M=\log\widehat Z_{\mathrm{FR},M}
-\log\widehat Z_{\mathrm{RR}_{R},M}$, so negative values favor
RR\textsubscript{R}. ``Change'' is the design-robustness value minus its
baseline counterpart. MC SD is the root-sum-square of the two models'
across-seed standard deviations. D1 and D2 alter the scoring partition, so
absolute log-evidence levels are not comparable with the baseline; the
within-target difference and its sign are the primary comparisons, and changes
in magnitude are descriptive. D2 also changes $K\delta$ from 102 to 202.
\end{minipage}
\end{table}

Table~\ref{tab:rrR_design_robust_evidence} shows
that $\widehat\Delta_M$ remains negative in all 12 comparisons. D1 makes the
difference more negative at every $M$, whereas D2 and D3 reduce its absolute
magnitude without reversing its sign. The smallest gap is $-1.99$ for D2 at
$M=1$, compared with an MC SD of 0.13. Over the grid
$M\in\{1,2,10,50\}$, the evidence-maximizing value for FR is $M=10$ under all
three designs. For RR\textsubscript{R}, it is $M=50$ under D1 and D2 and
$M=10$ under D3. In every case, the evidence-maximizing value of $M$ is
numerically well separated from the second-highest value.

\newpage
\textbf{Appendix F: Notation list}

\vspace{3mm}
\footnotesize
\renewcommand{\arraystretch}{0.92}
\begin{tabular}{@{}
 >{\raggedright\arraybackslash}p{0.30\textwidth}
 >{\raggedright\arraybackslash}p{0.60\textwidth}@{}}

\textbf{Models, data, and parameters} & \\
$A$, $E$, $\mathbf y$ & Structural model, empirical reference model, and observed data \\
$\theta_A$, $\theta_E$ & Structural- and empirical-model parameter vectors \\
$\mathcal A_A$, $\pi_A$ & Structural admissibility set and prior conditioned on that set
\\[2pt]

\textbf{Population moments and empirical target} & \\
$g_{A,i}$, $g_{E,i}$ & Structural and empirical mappings to population moment $i$ \\
$\mathcal T_E(\mathbf y)$ & Fixed empirical target $\{(\mathcal B_i,\mathbf n_i)\}_{i=1}^I$ \\
$\mathcal B_i$, $B_{k,i}$, $K$ & Partition for population moment $i$, its $k$th cell, and the number of cells \\
$n_{k,i}$, $c_{k,i}$ & Effective empirical and model-implied counts in cell $k$ \\
$\widehat{\mathbf p}_i$ & Empirical cell-frequency vector before count conversion \\
$N_{\mathrm{eff}}$, $N_{\mathrm{MC}}$ & Target information scale and number of empirical-model posterior draws
\\[2pt]

\textbf{Structural ensemble and DM score} & \\
$M$, $\Theta_{A,M}$ & Ensemble size and ensemble $(\theta_A^1,\ldots,\theta_A^M)$ \\
$\mathcal M$, $M_A^\star$, $\widehat M_A$ & Prespecified grid of $M$, evidence-maximizing value, and its SMC estimate \\
$Q_M$ & Equally weighted distribution of the structural ensemble \\
$p_{A,k,i}$, $\mathbf p_{A,i}$ & Prior-predictive probability of cell $k$ for population moment $i$, and its vector \\
$\mathbf p_i$, $\boldsymbol\alpha_i$ & Latent cell probabilities and structurally updated Dirichlet concentration \\
$\delta$ & Symmetric Dirichlet base concentration \\
$L_{i,M}$, $L_M$ & Moment-specific DM scores and their product
\\[2pt]

\textbf{Posterior and evidence} & \\
$\Pi_{A,M}^{0}$, $\Pi_{A,M}$ & Product prior and generalized posterior on the ensemble \\
$Z_{A,M}$, $\Delta_M(A_1,A_2)$ & Joint generalized evidence and common-$M$ log-evidence difference \\
$D_{\mathrm{KL}}$ & Kullback--Leibler divergence
\\[2pt]

\textbf{Annealed SMC} & \\
$P$, $\{\beta_t\}_{t=0}^T$ & Number of particles and annealing schedule \\
$\widetilde w_t^p$, $\mathrm{ESS}$ & Incremental weight and effective sample size \\
$\widehat Z_{A,M}$ & Annealed-SMC estimator of the generalized evidence
\end{tabular}

\vspace{4pt}
\noindent\emph{Notes:} Indices $i=1,\ldots,I$, $k=1,\ldots,K$, and
$j=1,\ldots,M$ denote population moments, cells, and ensemble members,
respectively. Table~1 of the main paper classifies the principal
quantities as target-defining, computational, or diagnostic.

%% file: appendix_rr_posterior_tables.tex
\begin{table}[H]
\centering
\captionsetup{font=footnotesize}
\caption{Posterior inference and generalized evidence for the RR\textsubscript{g} model under the baseline empirical target ($N_{\mathrm{MC}}=N_{\mathrm{eff}}=50{,}000$)}
\label{tab:posterior_rr_data}
\scriptsize
\setlength{\tabcolsep}{2.0pt}
\renewcommand{\arraystretch}{0.72}
\begin{adjustbox}{max width=\textwidth}
\begin{tabular}{llccccccc}
\hline
Parameter
& Prior mean
& $M=1$
& $M=2$
& $M=5$
& $M=10$
& $M=20$
& $M=30$
& $M=50$ \\
\hline
$\ln\gamma$ (\%)
& 0.500
& 0.584 & 0.600 & 0.612 & 0.623 & 0.631 & 0.635 & 0.637 \\
&
& [0.237, 0.838] & [0.251, 0.838] & [0.256, 0.837] & [0.268, 0.844] & [0.278, 0.847] & [0.286, 0.851] & [0.294, 0.856] \\

$\ln\pi^*$ (\%)
& 1.000
& 1.015 & 1.017 & 1.015 & 1.016 & 1.016 & 1.016 & 1.015 \\
&
& [0.730, 1.307] & [0.739, 1.303] & [0.728, 1.312] & [0.705, 1.328] & [0.691, 1.346] & [0.669, 1.363] & [0.642, 1.386] \\

$\ln r^*$ (\%)
& 0.500
& 0.499 & 0.501 & 0.502 & 0.504 & 0.503 & 0.504 & 0.504 \\
&
& [0.296, 0.747] & [0.297, 0.762] & [0.295, 0.754] & [0.296, 0.761] & [0.294, 0.761] & [0.292, 0.765] & [0.292, 0.771] \\

$\kappa$
& 0.300
& 0.282 & 0.233 & 0.218 & 0.215 & 0.214 & 0.214 & 0.214 \\
&
& [0.117, 0.460] & [0.103, 0.454] & [0.095, 0.440] & [0.086, 0.440] & [0.078, 0.448] & [0.074, 0.449] & [0.070, 0.455] \\

$\tau$
& 2.000
& 2.193 & 2.297 & 2.406 & 2.457 & 2.486 & 2.497 & 2.500 \\
&
& [1.677, 2.817] & [1.715, 2.973] & [1.757, 3.079] & [1.772, 3.138] & [1.756, 3.173] & [1.755, 3.204] & [1.757, 3.227] \\

$\psi_1$
& 1.500
& 1.477 & 1.465 & 1.459 & 1.458 & 1.456 & 1.455 & 1.454 \\
&
& [1.251, 1.720] & [1.230, 1.722] & [1.226, 1.715] & [1.228, 1.704] & [1.228, 1.708] & [1.224, 1.709] & [1.225, 1.706] \\

$\psi_2$
& 0.125
& 0.118 & 0.113 & 0.113 & 0.112 & 0.112 & 0.112 & 0.112 \\
&
& [0.047, 0.230] & [0.043, 0.226] & [0.043, 0.219] & [0.042, 0.219] & [0.042, 0.218] & [0.041, 0.217] & [0.041, 0.217] \\

$\rho_R$
& 0.500
& 0.651 & 0.721 & 0.768 & 0.785 & 0.793 & 0.795 & 0.797 \\
&
& [0.351, 0.864] & [0.346, 0.871] & [0.387, 0.877] & [0.393, 0.880] & [0.399, 0.884] & [0.401, 0.887] & [0.389, 0.890] \\

$\rho_z$
& 0.300
& 0.301 & 0.301 & 0.302 & 0.301 & 0.301 & 0.301 & 0.302 \\
&
& [0.208, 0.405] & [0.206, 0.404] & [0.209, 0.406] & [0.208, 0.404] & [0.208, 0.405] & [0.207, 0.405] & [0.207, 0.407] \\

$\sigma_R$
& 0.251
& 0.258 & 0.254 & 0.250 & 0.251 & 0.250 & 0.251 & 0.251 \\
&
& [0.101, 0.400] & [0.110, 0.404] & [0.105, 0.398] & [0.106, 0.400] & [0.103, 0.402] & [0.103, 0.401] & [0.102, 0.403] \\

$\sigma_z$
& 0.875
& 0.888 & 0.891 & 0.891 & 0.891 & 0.894 & 0.895 & 0.897 \\
&
& [0.791, 0.979] & [0.798, 0.985] & [0.797, 0.981] & [0.798, 0.986] & [0.798, 0.991] & [0.798, 0.995] & [0.798, 0.999] \\

\hline
$\log \widehat Z_{\mathrm{RR}_{g},M}$
& --
& -15217.08 & -15223.05 & -15257.03 & -15343.92 & -15580.67 & -15867.61 & -16529.23 \\
\hline
\end{tabular}
\end{adjustbox}
\end{table}

\noindent
\begin{minipage}{0.96\textwidth}
\scriptsize
\textit{Notes to Tables E.1--E.3.} Posterior means are reported in the first
row and 95\% credible intervals in the second row. The posterior summaries
pool the exchangeable structural-parameter draws from the final SMC particles
across the two independent runs used at the reported particle count. The
parameters $\ln\gamma$, $\ln\pi^*$, and $\ln r^*$ are expressed in percentage
terms. For the shock standard deviations, the summaries refer to
$\sigma_j=|s_j|$ as defined in Table~4 of the main paper. The final row
reports the estimated log joint generalized evidence. The RR\textsubscript{z}
and RR\textsubscript{R} specifications use model-specific annealing schedules.
\end{minipage}

\begin{table}[H]
\centering
\captionsetup{font=footnotesize}
\caption{Posterior inference and generalized evidence for the RR\textsubscript{z} model under the baseline empirical target ($N_{\mathrm{MC}}=N_{\mathrm{eff}}=50{,}000$)}
\label{tab:posterior_rrz_data}
\scriptsize
\setlength{\tabcolsep}{2.0pt}
\renewcommand{\arraystretch}{0.72}
\begin{adjustbox}{max width=\textwidth}
\begin{tabular}{llcccc}
\hline
Parameter
& Prior mean
& $M=1$
& $M=2$
& $M=10$
& $M=50$ \\
\hline
$\ln\gamma$ (\%)
& 0.500
& 0.595 & 0.606 & 0.628 & 0.643 \\
&
& [0.244, 0.835] & [0.251, 0.837] & [0.270, 0.847] & [0.298, 0.858] \\

$\ln\pi^*$ (\%)
& 1.000
& 1.022 & 1.019 & 1.017 & 1.017 \\
&
& [0.722, 1.306] & [0.721, 1.301] & [0.712, 1.323] & [0.645, 1.385] \\

$\ln r^*$ (\%)
& 0.500
& 0.503 & 0.498 & 0.499 & 0.499 \\
&
& [0.295, 0.758] & [0.291, 0.740] & [0.292, 0.749] & [0.289, 0.761] \\

$\kappa$
& 0.300
& 0.335 & 0.339 & 0.339 & 0.339 \\
&
& [0.201, 0.488] & [0.207, 0.496] & [0.198, 0.496] & [0.190, 0.502] \\

$\tau$
& 2.000
& 1.950 & 1.947 & 1.948 & 1.940 \\
&
& [1.490, 2.468] & [1.495, 2.431] & [1.502, 2.461] & [1.477, 2.462] \\

$\psi_1$
& 1.500
& 1.360 & 1.358 & 1.362 & 1.356 \\
&
& [1.167, 1.583] & [1.160, 1.575] & [1.161, 1.586] & [1.148, 1.591] \\

$\psi_2$
& 0.125
& 0.229 & 0.230 & 0.229 & 0.228 \\
&
& [0.140, 0.346] & [0.138, 0.349] & [0.134, 0.350] & [0.120, 0.360] \\

$\rho_R$
& 0.500
& 0.427 & 0.428 & 0.433 & 0.448 \\
&
& [0.248, 0.609] & [0.248, 0.613] & [0.241, 0.649] & [0.236, 0.730] \\

$\rho_g$
& 0.800
& 0.928 & 0.928 & 0.929 & 0.927 \\
&
& [0.898, 0.955] & [0.895, 0.956] & [0.893, 0.961] & [0.877, 0.962] \\

$\sigma_R$
& 0.251
& 0.283 & 0.280 & 0.280 & 0.283 \\
&
& [0.135, 0.433] & [0.116, 0.439] & [0.108, 0.457] & [0.092, 0.490] \\

$\sigma_g$
& 0.630
& 0.706 & 0.713 & 0.731 & 0.750 \\
&
& [0.575, 0.851] & [0.558, 0.863] & [0.546, 0.959] & [0.546, 1.008] \\

\hline
$\log \widehat Z_{\mathrm{RR}_{z},M}$
& --
& -15200.37 & -15188.29 & -15157.50 & -15608.21 \\
\hline
\end{tabular}
\end{adjustbox}
\end{table}

\begin{table}[H]
\centering
\captionsetup{font=footnotesize}
\caption{Posterior inference and generalized evidence for the RR\textsubscript{R} model under the baseline empirical target ($N_{\mathrm{MC}}=N_{\mathrm{eff}}=50{,}000$)}
\label{tab:posterior_rrR_data}
\scriptsize
\setlength{\tabcolsep}{2.0pt}
\renewcommand{\arraystretch}{0.72}
\begin{adjustbox}{max width=\textwidth}
\begin{tabular}{llcccc}
\hline
Parameter
& Prior mean
& $M=1$
& $M=2$
& $M=10$
& $M=50$ \\
\hline
$\ln\gamma$ (\%)
& 0.500
& 0.599 & 0.603 & 0.626 & 0.642 \\
&
& [0.248, 0.845] & [0.252, 0.836] & [0.269, 0.844] & [0.297, 0.859] \\

$\ln\pi^*$ (\%)
& 1.000
& 1.018 & 1.016 & 1.018 & 1.018 \\
&
& [0.736, 1.287] & [0.727, 1.294] & [0.710, 1.323] & [0.644, 1.390] \\

$\ln r^*$ (\%)
& 0.500
& 0.500 & 0.498 & 0.499 & 0.500 \\
&
& [0.288, 0.735] & [0.297, 0.737] & [0.292, 0.752] & [0.288, 0.765] \\

$\kappa$
& 0.300
& 0.326 & 0.329 & 0.331 & 0.335 \\
&
& [0.192, 0.480] & [0.193, 0.479] & [0.197, 0.484] & [0.197, 0.489] \\

$\tau$
& 2.000
& 2.002 & 2.006 & 2.001 & 2.006 \\
&
& [1.552, 2.526] & [1.544, 2.532] & [1.532, 2.523] & [1.543, 2.527] \\

$\psi_1$
& 1.500
& 1.358 & 1.360 & 1.355 & 1.342 \\
&
& [1.154, 1.580] & [1.154, 1.586] & [1.151, 1.579] & [1.132, 1.576] \\

$\psi_2$
& 0.125
& 0.236 & 0.239 & 0.246 & 0.255 \\
&
& [0.136, 0.362] & [0.137, 0.368] & [0.143, 0.376] & [0.145, 0.394] \\

$\rho_R$
& 0.500
& 0.461 & 0.463 & 0.460 & 0.467 \\
&
& [0.279, 0.643] & [0.280, 0.646] & [0.277, 0.653] & [0.269, 0.672] \\

$\rho_g$
& 0.800
& 0.931 & 0.931 & 0.931 & 0.929 \\
&
& [0.901, 0.957] & [0.903, 0.958] & [0.900, 0.957] & [0.891, 0.958] \\

$\rho_z$
& 0.300
& 0.302 & 0.302 & 0.311 & 0.319 \\
&
& [0.207, 0.420] & [0.203, 0.424] & [0.192, 0.448] & [0.177, 0.472] \\

$\sigma_g$
& 0.630
& 0.693 & 0.690 & 0.684 & 0.688 \\
&
& [0.538, 0.858] & [0.530, 0.850] & [0.534, 0.838] & [0.534, 0.851] \\

$\sigma_z$
& 0.875
& 0.866 & 0.859 & 0.855 & 0.850 \\
&
& [0.748, 0.982] & [0.748, 0.982] & [0.747, 0.976] & [0.742, 0.971] \\

\hline
$\log \widehat Z_{\mathrm{RR}_{R},M}$
& --
& -15201.88 & -15189.45 & -15131.00 & -15311.87 \\
\hline
\end{tabular}
\end{adjustbox}
\end{table}